\documentclass[fleqn,twocolumn,superscriptaddress]{revtex4}
\usepackage[utf8x]{inputenc}
 \usepackage{epsfig}
\usepackage{amssymb}
 \usepackage{amsthm}
 \usepackage{color}
\usepackage{amsmath}
\usepackage{empheq}

\begin{document}

\title{Coarsening dynamics in one dimension: the phase diffusion equation\\and its numerical implementation}



\author{Matteo Nicoli}
\affiliation{Physique de la Mati\`ere Condens\'ee, \'Ecole Polytechnique, CNRS,  Palaiseau, F-91128 France}
\altaffiliation[Present address: ]{Center for Interdisciplinary Research on Complex Systems,
Department of Physics, Northeastern University, Boston, Massachusetts 02115, USA.}
\author{Chaouqi Misbah}
\affiliation{Univ. Grenoble 1 / CNRS, LIPhy UMR 5588, Grenoble, F-38041, France}
\author{Paolo Politi}
\affiliation{Istituto dei Sistemi Complessi, Consiglio Nazionale Delle Ricerche, via Madonna Del Piano 10, I-50019 Sesto Fiorentino , Italy}

\begin{abstract}
Many nonlinear partial differential equations (PDEs) display a coarsening dynamics, i.e., an emerging pattern whose typical
length scale $L$ increases with time.
The so-called coarsening exponent $n$ characterizes the time dependence of the scale of
the pattern, $L(t)\approx t^n$, and
coarsening dynamics can be described by a diffusion equation for the phase
of the pattern.
By means of a multiscale analysis
we are able to find the  analytical expression of such diffusion equations.
Here, we propose
a recipe to implement numerically the determination of $D(\lambda)$, the phase diffusion coefficient,
as a function of the wavelength $\lambda$ of the base steady state $u_0(x)$.
$D$ carries  all  information about
coarsening dynamics and, through the relation  $|D(L)| \simeq L^2 /t$,
it allows us to determine the coarsening exponent. The main conceptual 
message is that the coarsening exponent is determined 
without solving a time-dependent equation, but only by inspecting the periodic steady-state solutions. 
This provides a much faster strategy than a forward time-dependent calculation.
We discuss our method for several different PDEs, both conserved and not conserved.
\end{abstract}

\maketitle


\section{Introduction}
\label{sec_intro}

In this paper we will focus on the nonlinear dynamics of one-dimensional (1D) partial differential
equations (PDEs), having the form~\cite{Politi:2006}
\begin{equation}
\partial_t u = {\cal N}[u] ,
\label{nle}
\end{equation}
where ${\cal N}$ is a general nonlinear operator, not depending explicitly on time $t$ and space $x$
and whose trivial solution $u\equiv 0$ has a linear stability spectrum 
whose modes of sufficiently small wave vector are unstable. More precisely, setting
$u(x,t) = \bar u \exp(iqx + \omega t)$, and keeping only terms that  are linear in
$\bar u$, one obtains a dispersion relation  relating $\omega$ to $q$. Two well-known dispersion relations~\cite{Politi:2006} are,
\begin{subequations}
\begin{align}
& \omega (q) =  1 - q^2   & & \mbox{nonconserved models}, &
\label{w_nc} \\[5pt]
& \omega (q) = q^2 - q^4  & &  \mbox{conserved models}. &
\label{w_con}
\end{align}
\label{eq_omega}
\end{subequations}
In both cases   $\omega (q)>0$ for $q<1$. In other words, the trivial solution $u=0$ is unstable for $q<1$. 
This means that modes with arbitrary small wave numbers are unstable, and therefore, in principle, spatial structures 
with large wavelength can develop, so that the notion of coarsening makes sense. This implies that  we exclude situations 
where there is a small $q$ cut-off, that is, when there is a minimal wave number below which the trivial solution is stable.
A typical example in this category is the Swift-Hohenberg equation \cite{Politi:2006}, 
for which the dispersion relation  takes the form  $\omega(q)=\alpha +(q-1)^2$, with $\alpha$ a real parameter. This dispersion relation means that if $\alpha<0$, then there is a minimal wave vector $q_{min}=1-\sqrt{\alpha}$ below which the trivial solution is stable. We do not expect in this case a structure with a wave number smaller than $q_{min}$ to take place (no perpetual coarsening is {\sl a priori} expected). While this statement complies with intuition, we must keep in 
mind that the stability evoked above follows from a linear analysis. Therefore, since the equations of interest are nonlinear, exceptions are not excluded~\cite{note1}.
Having in mind the occurrence of some exceptions, we will focus on systems having a linear  dispersion relation of the form
 (\ref{eq_omega}). This holds for many nonlinear PDEs.
The most well-known equations are the Ginzburg-Landau (GL), the Cahn-Hilliard (CH), the Kuramoto-Sivashinsky
(KS), and their variants or generalizations~\cite{Politi:2006}. Equations of this class will be introduced and discussed in the
next section.

Note that having  unstable modes for small $q$ as in (\ref{eq_omega}) is not a sufficient condition. A prominent example
is the KS equation, which shows
spatiotemporal chaos, with wavelengths of constant average size being continuously
destroyed and created.
This is not surprising: the analysis of the linear stability spectrum does not univocally determines
the nonlinear dynamics.

One important feature that we will discuss here is the determination   of the branch of periodic steady-state solutions, which acquire a special status in our investigation of coarsening dynamics. This is {\sl a priori} not obvious, since coarsening is a time-dependent process, 
but it has been shown before that stationary configurations of period $\lambda$, $u_0 (x)$,
play a major role in determining the type of nonlinear dynamics: for a certain class of nonlinear equations in one dimension, two of us
have proven~\cite{PRL} that coarsening is possible if and only if the wavelength $\lambda$ is an increasing
function of the amplitude $A$ of the solution $u_0(x)$, $\lambda'(A)>0$.
The presence of a maximum in the curve $\lambda(A)$ signals the phenomenon of
interrupted coarsening, while a decreasing $\lambda (A)$ signals no coarsening at all. Finally,
for some equations such as  KS, $\lambda(A)$ is not a single-value function and the curve
displays a so-called turning point.

All these examples show that steady states contain important pieces of information. However,
how is it possible that dynamics depends on stationary properties only?
The answer to this question  is contained in the concept of phase instability. For example, in one dimension, a steady-state periodic solution $u_0(x)$ has a wavelength (or wave number $q=2\pi/\lambda$) $\lambda$ such that $u_0(x+\lambda)=u_0(x)$. In this case the phase $\phi$ of the pattern is $\phi=qx$. However, a perturbation of the wavelength of the pattern (or equivalently a perturbation of the phase $\phi$) may reveal instability (called phase instability), meaning that the wavelength will evolve in time, and if this is the case for any $q$, one may expect coarsening. A global shift of a periodic pattern in the $x$ direction by a constant value $x_0$ is a also a solution, i.e., $u_0(x+x_0)=u_0(x)$, owing to translational invariance (also called a Goldstone mode). This constant shift corresponds to a phase perturbation of infinite wavelength, and since it is a stationary solution, it has an infinite relaxation time towards the original solution $u_0(x)$. In other words, a large wavelength perturbation of the pattern is expected to have a large time scale. This entails that the most dangerous perturbations are those of long wavelengths of the phase of the pattern. To make this notion explicit, we introduce a small parameter $\epsilon$ expressing the fact that the phase modes of interest have a small spatial gradient, on the scale of the wavelength of the pattern. An appropriate way to deal with this question is to introduce a multiscale analysis: the fast spatial scale $x$ (where the periodic profile of pattern  varies on a scale of order unity) and a slow scale $X$ (expressing the long wavelength perturbation of the phase of the pattern). The slow spatial scale, implies also a slow time scale (which can also be inferred from the above discussion about large relaxation times) $T$. It turns out that (see later) $X=\epsilon x$ and $T=\epsilon ^2 t$.

 If $\phi$ designates the actual phase of a pattern (not necessarily a periodic one, but some perturbed pattern), it will depend on space and time, e.g., because
$u_0(x)$ is perturbed, and acquires a slow spatial and time dependence (i.e., it depends on $X$ and $T$), which can be described
by a phase diffusion equation (see next section), having the form
\begin{equation}
\partial_T \psi = D(q) \partial_{XX} \psi ,
\end{equation}
where $\psi,X,T$ are suitably rescaled versions of $\phi,x,t$, and the phase diffusion
coefficient $D(q)$ only depends on properties of the steady-state solution of wavelength
$\lambda=2\pi/q$.
A negative $D(q)$ signals a phase instability, i.e., coarsening. Other scenarios are possible~\cite{note2}.
Even if for several nonlinear equations it is feasible  to obtain the analytical form of
$D(q)$ in the limit of large $\lambda$, in the generic case
it is not possible
and we must determine it numerically, by computing the steady-state solution $u_0(x)$, from which we can state if the pattern is stable (no coarsening), or unstable (coarsening). We will then show how the coarsening exponent (in the temporal power law) can be extracted from these considerations (i.e. without any time integration of the equations).

In this paper we start by briefly  describing the multiscale method and apply it to several 1D PDEs,
therefore extracting the quantities we need to determine $D(q)$ numerically. We will see that we need two
functions: (i)~$u_0(x)$, the stationary solution, which satisfies
the equation ${\cal N}[u_0]=0$, see Eq.~(\ref{nle}), and
(ii)~$v_0(x)$, solution of the equation ${\cal L}^\dag [v_0(x)]=0$, where
${\cal L}$ is the Frech\'et derivative of ${\cal N}$ and ${\cal L}^\dag$ is its adjoint operator.
We will show how it is possible to implement the numerical determination of such functions,
therefore the determination of $D(q)$ and, finally, how we can get the coarsening exponent $n$,
describing the time dependence of the typical scale, $L \approx t^n$, when coarsening is present.
Particular attention will be devoted to the conserved Kuramoto-Sivashinsky (cKS) equation, because
in such a case ${\cal L}$ is not self-adjoint and special care must be devoted to the determination
of $v_0(x)$.

\section{From the 1D nonlinear equation to the phase diffusion equation}
\label{sec_eqs}

In this section we introduce all nonlinear equations we are going to discuss, we give a very
brief sketch of the multiscale approach used to extract the phase diffusion equation, and we finally provide the expression of the phase
diffusion coefficient $D$. In some cases, we just report existing results,
but in the case of the cKS equation, we provide the original derivation of
$D$, which will be seen to deserve special attention.

\subsection{The models}
\label{sec_GL}

Let us start by listing all equations of interest, the nonconserved first:
\begin{subequations}
\label{eq_nc}
\begin{align}
\label{GL_eq}
&\partial_t u = u - u^3 +\partial_x^2 u & & \mbox{GL~equation}, & \\[5pt]
\label{nonc_alpha_eq}
&\partial_t u = \dfrac{u}{\left(1+ u^2 \right)^\alpha}  + \partial_x^2 u , & &\mbox{$\alpha$ models}.&
\end{align}
\end{subequations}
then their conservative counterparts:
\begin{subequations}
\label{eq_con}
\begin{align}
\label{CH_eq}
&\hskip -10pt \partial_t u = -\partial_x^2\left[ u - u^3 +\partial_x^2 u\right] & & \mbox{CH equation,}& \\[5pt]
\label{c_alpha_eq}
&\hskip -10pt\partial_t u = -\partial_x^2\left[ \dfrac{u}{\left(1+ u^2 \right)^\alpha}  + \partial_x^2 u\right] & &\mbox{c$\alpha$ models}.&
\end{align}
\end{subequations}
and, finally, the conserved Kuramoto-Sivashinsky equation
\begin{equation}
\label{cKS_eq}
\partial_t u = -\partial^2_x [ u - \tau \partial_x u + \partial_x^2 u -\left( \partial_x u\right)^2 ]\quad   \mbox{cKS~equation}.
\end{equation}

GL and CH are classical equations~\cite{Bray} describing, e.g., phase separation processes for a
scalar order parameter, either conserved (CH) or nonconserved (GL).
The c$\alpha$ model for $\alpha=1$ first  appeared in problems of crystal growth~\cite{PPJV}
and starting from that specific conserved model it has been natural to extend it~\cite{ATPP}
to $\alpha\ne 1$ and to the nonconserved case.
Finally, the cKS equation appeared in the study of sand-ripple dynamics~\cite{Csahok:2000},
then in step-bunching phenomena~\cite{Gillet}. For $\tau=0$ it also appears  in the
study of wandering crystalline steps~\cite{FV,CKS}.

Equations~(\ref{eq_nc}) have the general form $\partial_t u = A(u) + \partial_x^2 u$
and their linear dispersion relation has the form (\ref{w_nc}).
All conservative equations, (\ref{eq_con}) and (\ref{cKS_eq}), have the form
$\partial_t u = -\partial_x^2\left(\cdots \right)$
and their linear dispersion relation is like (\ref{w_con}). For Eq.~(\protect\ref{cKS_eq}), that is true
for $\tau=0$ only. However, the $\tau$ term can be canceled out with a tilt transformation:
$u \to u +\tau x/2$. We keep it here for sake of generality.

All previous models, conserved and nonconserved,
display perpetual coarsening, because, as analyzed in the next subsection,
they have a branch $u_0(x)$ of steady states whose $\lambda(A)$ characteristic is an
increasing, diverging function.

\subsection{The multiscale method and the phase diffusion equation}

We are now going to discuss the multiscale method, with a focus on the cKS eq.
This step has two motivations. First, the cKS equation  is a completely non trivial equation,
whose nonlinear dynamics has been studied only through direct numerical simulations or through
a simplified particle model. We are not aware of analytical nonlinear studies.
Second, its treatment with the multiscale approach is representative of the method, when
the Frech\'et derivative is not self-adjoint.

The basic idea is to perturb a periodic steady state
$u_0(x,\lambda)=u_0(x+\lambda,\lambda)$.
First, we rescale $x$ by introducing the phase $\phi=qx$, with $q=2\pi/\lambda$,
so that $u_0=u_0(\phi,q)$. Next, we 
introduce slow space and time variables,
\begin{equation}
X= \epsilon x, \qquad T = \epsilon^2 t,
\label{slow}
\end{equation}
and perturb $u_0$ so that $q$ is now dependent on $X$ and $T$. While $x$ is
the fast space variable, 
no fast time variable exists, since the base state $u_0(x)$ is a stationary state. 
The final step is to perform a
 multiple scale expansion method~\cite{multiscale}.
Identifying the slow phase $\psi(X,T)=\epsilon \phi(x,t)$, 
we can finally rewrite the space and time derivatives,
\begin{subequations}
\label{derivatives}
\begin{align}
\partial_x &= q\partial_\phi +\epsilon \partial_X =q\partial_\phi +\epsilon\psi_{XX}\partial_q ,\\[5pt]
\partial_t &= \epsilon \psi_T \partial_\phi .
\end{align}
\end{subequations}

This is enough, along with the $\epsilon$ expansion of the solution,
$u = u_0 + \epsilon u_1$, to perform a correct perturbative analysis of Eq. (\ref{nle}).
Using Eqs.~\eqref{slow}--\eqref{derivatives}, the general nonlinear equation (\ref{nle}) rewrites as
\begin{equation}
\label{ms_eq}
\begin{split}
\epsilon (\partial_\phi u_0) \partial_T \psi =&
( {\cal N}_0 + \epsilon {\cal N}_1 ) [u_0 + \epsilon u_1 ] \\[5pt]
=& {\cal N}_0 [u_0] + \epsilon {\cal L}_0 [u_1] + \epsilon {\cal N}_1 [u_0] ,
\end{split}
\end{equation}
where ${\cal L}_0$ is the functional, Frech\'et derivative of ${\cal N}_0$.
Since ${\cal N}_0 [u_0]$ trivially vanishes, we are left with the first-order terms,
which can be rearranged, so that the function $u_0$ appears in the source term, $g$, of the linear
differential equation for $u_1$,
\begin{equation}
{\cal L}_0 [u_1] = (\partial_\phi u_0) \partial_T \psi - {\cal N}_1 [u_0] \equiv g .
\end{equation}
According to the Fredholm alternative theorem~\cite{Fredholm}, $g$ must be orthogonal to the kernel
of the adjoint operator, ${\cal L}_0^\dag$. In conclusion, if
\begin{equation}
{\cal L}_0^\dag [v_0] = 0 ,
\label{eq_v0}
\end{equation}
then
\begin{equation}
\langle v_0, (\partial_\phi u_0) \partial_T \psi - {\cal N}_1 [u_0] \rangle =0 .
\label{Fredholm}
\end{equation}

In the next subsection we are going to see that Eq.~(\ref{Fredholm}) can be recast  in the form
of a phase diffusion equation,
\begin{equation}
\partial_T \psi = D(q) \; \partial_{XX} \psi ,
\label{phase_diffusion}
\end{equation}
where the phase diffusion coefficient, $D(q)$ is a nontrivial functional of $u_0$ and $v_0$
\cite{nota}.

\begin{figure*}[t!]
\begin{center}
\begin{minipage}{0.495\textwidth}
\includegraphics[height = 6cm,clip=]{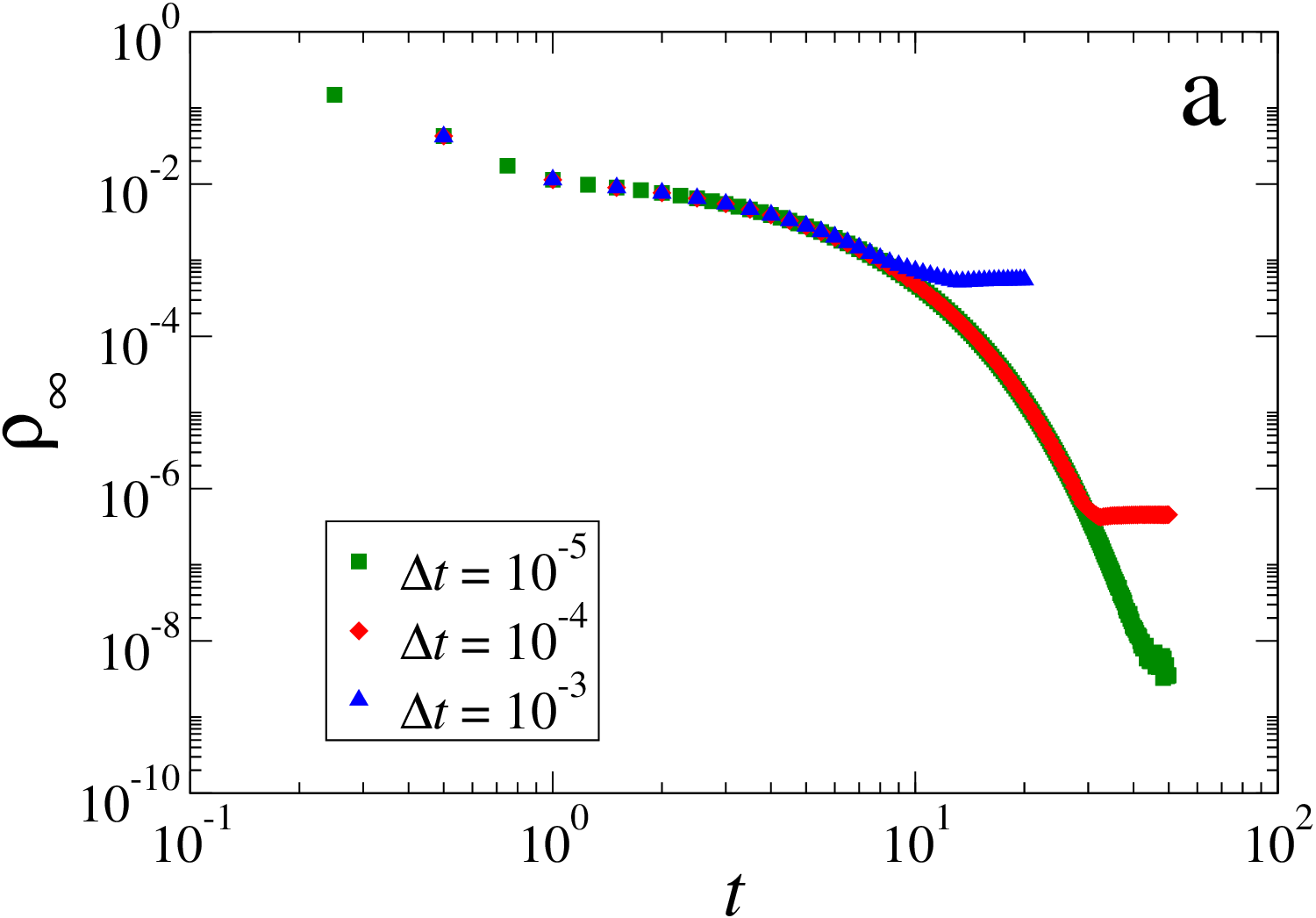}
\end{minipage}
\begin{minipage}{0.495\textwidth}
\includegraphics[height = 6.05cm,clip=]{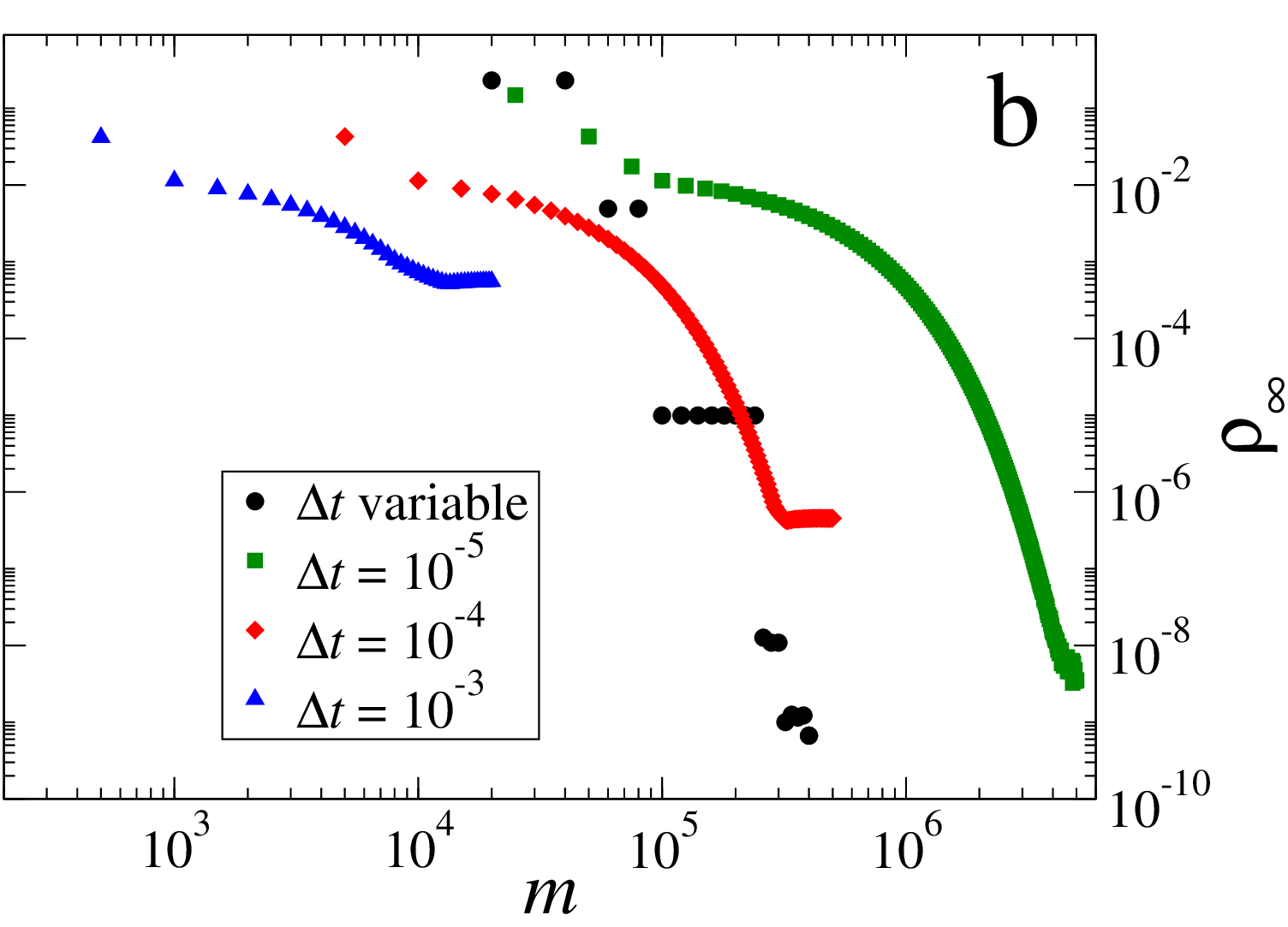}
\end{minipage}
\caption{(a) Evolution of the residue $\rho_\infty$ of the Cahn-Hilliard equation
\eqref{CH_eq} as function of time for different values of $\Delta t$ (reported in the legend).
The initial condition is a sinusoid discretized by $N=512$ points
with $\lambda = 12.8$,  corresponding  to $q = 0.49087$.
Note that for a fixed $\Delta t$ the residue saturates at an asymptotic value, which depends on the magnitude
of the time steps.
(b) Evolution of $\rho_\infty$ as function of the iteration step $m$ for the same simulation condition of the left figure.
The black points are the value of $\rho_\infty$ for a numerical
integration of Eq.~\eqref{CH_eq} with a variable time stepping. Starting from $\Delta t = 0.025$, the magnitude  of $\Delta t$
is reduced of a factor ten ($\delta t = 0.1$) every time the condition  $\Delta \rho_\infty < 10^{-4}$ is verified.
The numerical integration stops for  $\rho_\infty <10^{-9}$, value reached for $\Delta t = 2.5\times 10^{-6}$.
Note that the sudden improvements of the accuracy of the estimated solution $u_0$ are due to the
reduction of $\Delta t$.}
\label{fig_timestep}
\end{center}
\end{figure*}

\subsection{The phase diffusion coefficient}

The application of the multiscale method to Eqs. (\ref{eq_nc}) and (\ref{eq_con}) gives,
respectively
\begin{equation}
\label{Dred_GL}
D(q) = \frac{\partial_q\left\langle q \left( \partial_\phi u_0\right)^2\right\rangle}
	{\left\langle \left(\partial_\phi u_0 \right)^2\right\rangle}
\qquad
\mbox{nonconserved Eqs. (\ref{eq_nc}),}
\end{equation}
and
\begin{equation}
\label{Dred_CH}
D(q) =q^2 \dfrac{\partial_q\left\langle q \left( \partial_\phi u_0\right)^2\right\rangle}{\left\langle u_0^2\right\rangle}
\qquad
\mbox{conserved Eqs. (\ref{eq_con}).} 
\end{equation}

For the cKS equation (\ref{cKS_eq}), we give here below a few details for the derivation of $D(q)$.
In that case, the operators appearing in Eqs.~(\ref{ms_eq}-\ref{Fredholm}) are
\begin{widetext}
\begin{align}
& \mathcal{N}_0 \left[u_0\right] = -q^2\partial_\phi^2\left[\left(1 -\tau q\partial_\phi + q^2\partial_\phi^2\right)  u_0 
	-q^2 \left( \partial_\phi u_0 \right)^2\right],\label{N0} \\[5pt]
& \mathcal{L}_0\left[u_1\right] =  -q^2\partial_\phi^2\left[1-\tau q\partial_\phi + q^2\partial_\phi^2
	- 2 q^2 \left( \partial_\phi u_0 \right)\partial_\phi\right] u_1, \\[5pt]
& \mathcal{N}_1\left[u_0\right] = -  \psi_{XX} \left\{
q^2 \partial_\phi^2 \left[\left(2q\partial_q +1\right)\partial_\phi u_0 -\tau \partial_q u_0   -
	2  q\left(\partial_\phi u_0 \right) \partial_q u_0\right]\right\}, \\[5pt]
\label{eq_adjoint}
& \mathcal{L}_0^\dagger[v] = -q^2\left\{1 + \tau q\partial_\phi +q^2\partial_\phi^2
 	+2 q^2\left[\left(\partial_\phi^2 u_0 \right)+ \left(\partial_\phi u_0\right)
	\partial_\phi\right]\right\}\partial_\phi^2 v.
\end{align}
Finally, from Eq.~(\ref{Fredholm}) we can determine the diffusion coefficient,
\begin{equation}
\label{eq_D}
D(q)  = -q^2    \langle \partial_\phi^2 v_0, \left(2q\partial_q +1\right)\partial_\phi u_0
-\nu \partial_q u_0 - 2 q \left( \partial_q u_0 \right) \partial_\phi u_0\rangle /
\langle v_0,\partial_\phi u_0 \rangle .
\end{equation}
\end{widetext}

When ${\cal L}_0$ is self-adjoint, as for Eqs.~(\ref{eq_nc}), translational invariance immediately
provides the result $v_0 = \partial_\phi u_0$. For Eqs.~(\ref{eq_con}), ${\cal L}_0^\dag\ne{\cal L}_0$,
but Eq.~(\ref{eq_v0}) can be easily solved analytically and $v_0$ is still related to $u_0$
through some differential operator. In all these cases, covering Eqs.~\eqref{GL_eq}--\eqref{c_alpha_eq},
$D$ can be expressed as a function of $u_0$ and its derivatives only, see
Eqs.~\eqref{Dred_GL}--\eqref{Dred_CH}.
Finally, there are cases where ${\cal L}_0^\dag\ne{\cal L}_0$ and $v_0$ cannot be determined
analytically. The cKS equation is a typical example. In the next section we discuss the most
general implementation for a numerical determination of $u_0,v_0$ and $D$.

\begin{figure*}[t!]
\begin{center}
\begin{minipage}{0.495\textwidth}
\includegraphics[height = 6cm,clip=]{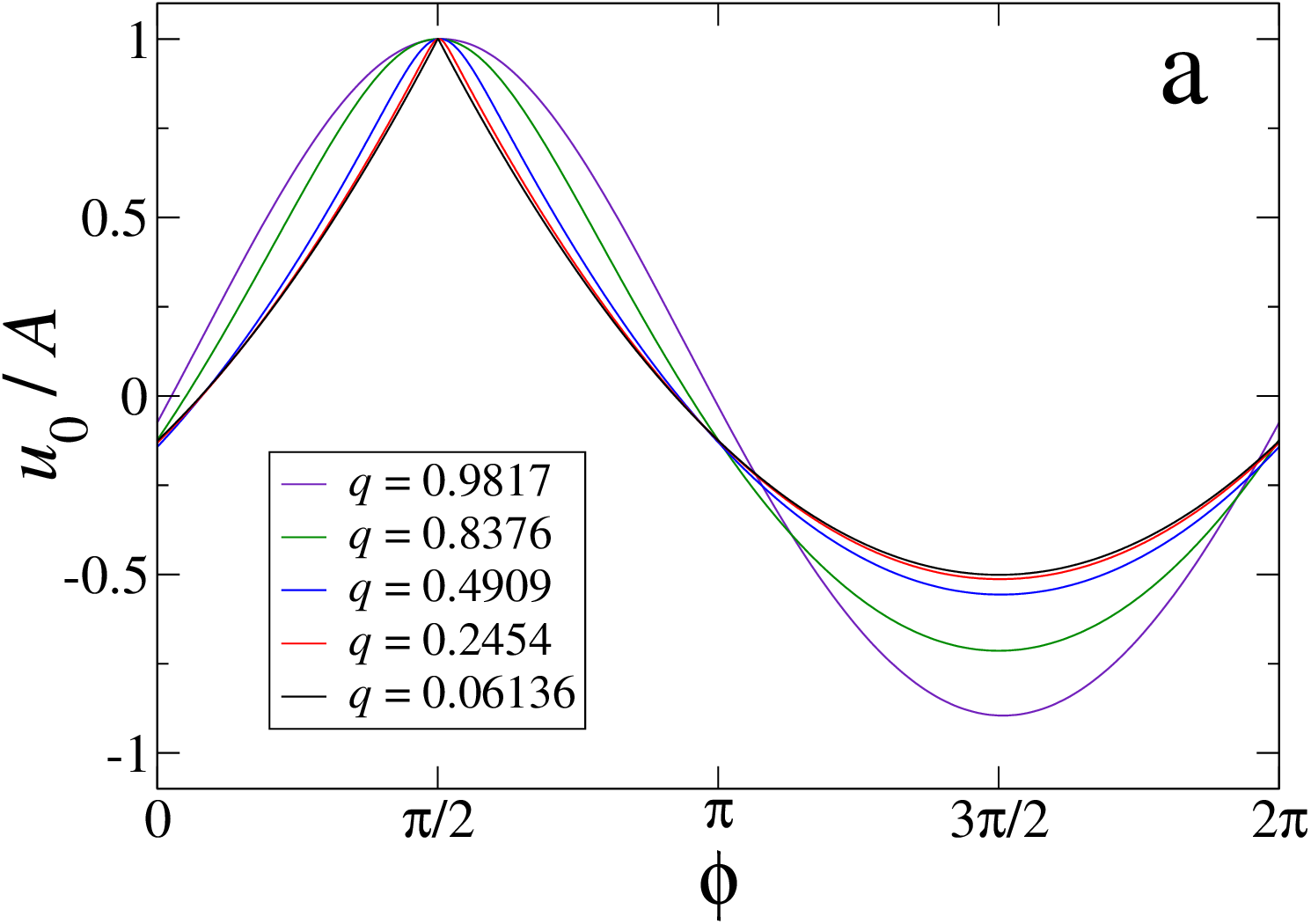}
\end{minipage}
\begin{minipage}{0.495\textwidth}
\includegraphics[height = 6cm,clip=]{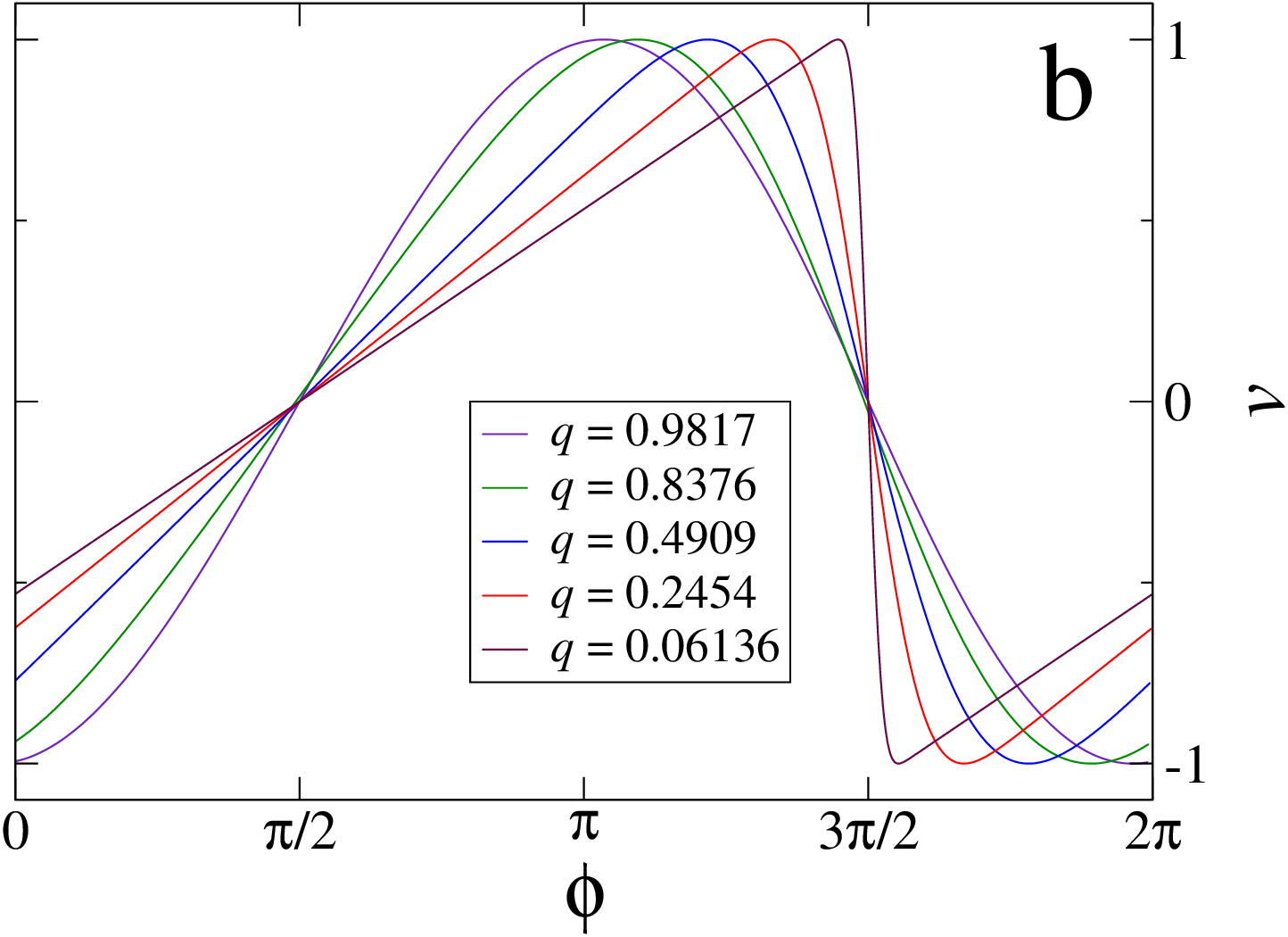}
\end{minipage}
\caption{(a)
Normalized stationary solutions $u_0$ of equation \eqref{cKS_phi_eq} with $\tau = 0$
for different values of $q$. The differential operators are discretized in an uniform grid with $\Delta \phi = 0.025$.
For visualization purposes we have chosen a normalization constant $A = \max |u(\phi)|$.
(b) Solutions of the adjoint problem $\mathcal{L}_0^\dagger[v_0]=0$, see Eq.~\eqref{eq_adjoint}, with $\tau = 0$.
As for the left figure, the spatial discretization is $\Delta \phi = 0.025$.}
\label{fig_u0_v}
\end{center}
\end{figure*}

\section{Numerical Implementation}
\label{sec_num}

The numerical procedure we have developed
to find the value of $D$ for a given $\lambda$ can be split in
three parts: \mbox{(i) The} determination of the stationary solution $u_0$  by employing 
a pseudospectral algorithm (see Appendix~\ref{ax_pseudo}) and, 
only if required, a  Newton-Raphson method to increase the accuracy of the output of our
pseudospectral code.
(ii) In the second step, the function $v_0$
is found by discretizing the adjoint ${\cal L}_0^\dag$ of the Frech\'et derivative of $\mathcal{N}_0$
with high order finite differences and finding the kernel of the
sparse matrix $M$ containing the coefficients of the discrete version of $\mathcal{L}^\dag_0$.
Clearly, we skip this step  when we are able to obtain  $v_0$ analytically.
(iii) Lastly, the value of $D(\lambda)$ is computed with a standard numerical integration after the evaluation of
the derivatives (in $\phi$ and $q$ space) of $u_0$ and, if necessary,  $v_0$ (only for the cKS equation).
These three steps are iterated for each $\lambda$ and, finally, the coarsening exponent is estimated
using the relation $|D|\sim L^2 / t$.

\subsection{Determination of the stationary solutions}

The time-dependent problem \eqref{nle} can be integrated very efficiently by using a pseudospectral
algorithm.
These  methods are widely used for the integration of nonlinear
PDEs because they combine the  estimation of spatial derivatives in Fourier space with
the  calculation of  nonlinear terms in real space, hence avoiding the computational overhead  of the convolution step
\cite{Trefethen_book,Boyd_book,Fornberg_book,Canuto_book}.
Pseudospectral  methods mainly work in Fourier space and their strength resides in their high accuracy for smooth solutions.

It is convenient to restate the nonlinear problem \eqref{nle} in $\phi \in [0,2\pi]$ space, by  replacing
the derivatives $\partial_x$ with $q\partial_\phi$. In this way, stationary solutions obtained at different
values of $q$ are mapped into the same interval so that they can be directly compared.

Every nonlinear PDE
such as Eq.~\eqref{nle} can be divided into a linear operator $\omega$, the dispersion relation, and a
nonlinear operator $\rm{N}$, such that in Fourier space  Eq.~\eqref{nle} reads
\begin{equation}
\label{nleFourier}
\partial_t u_k(t) = \omega_k u_k(t) + {\rm{N}} \left[ u(t) \right]_k,
\end{equation}
where ${\rm{N}}[u]_k $ is the Fourier transform of the nonlinear part of the PDE, and the discrete wave vectors $k$
belong to the  interval $[-N/2,N/2]$, with the same $\Delta k = 1$ for every $\lambda$ (here $N$ is the number of
points used to discretize the $\phi$ interval).
Equation~\eqref{nleFourier} is evaluated by using  the {\sl integrating factor} technique \cite{Trefethen_book,Canuto_book}
and time stepped with a fourth-order Adams-Bashforth-Moulton predictor-corrector scheme
(IFABM4) \cite{numerical_recipes}
(see Appendix~\ref{ax_pseudo} for the details of the numerical implementation).
For  a comparison between different pseudospectral algorithms  see
Ref.~\cite{Kassam:2005}  and references therein.

\begin{figure*}[t!]
\begin{center}
\includegraphics[width=0.6\textwidth,clip=]{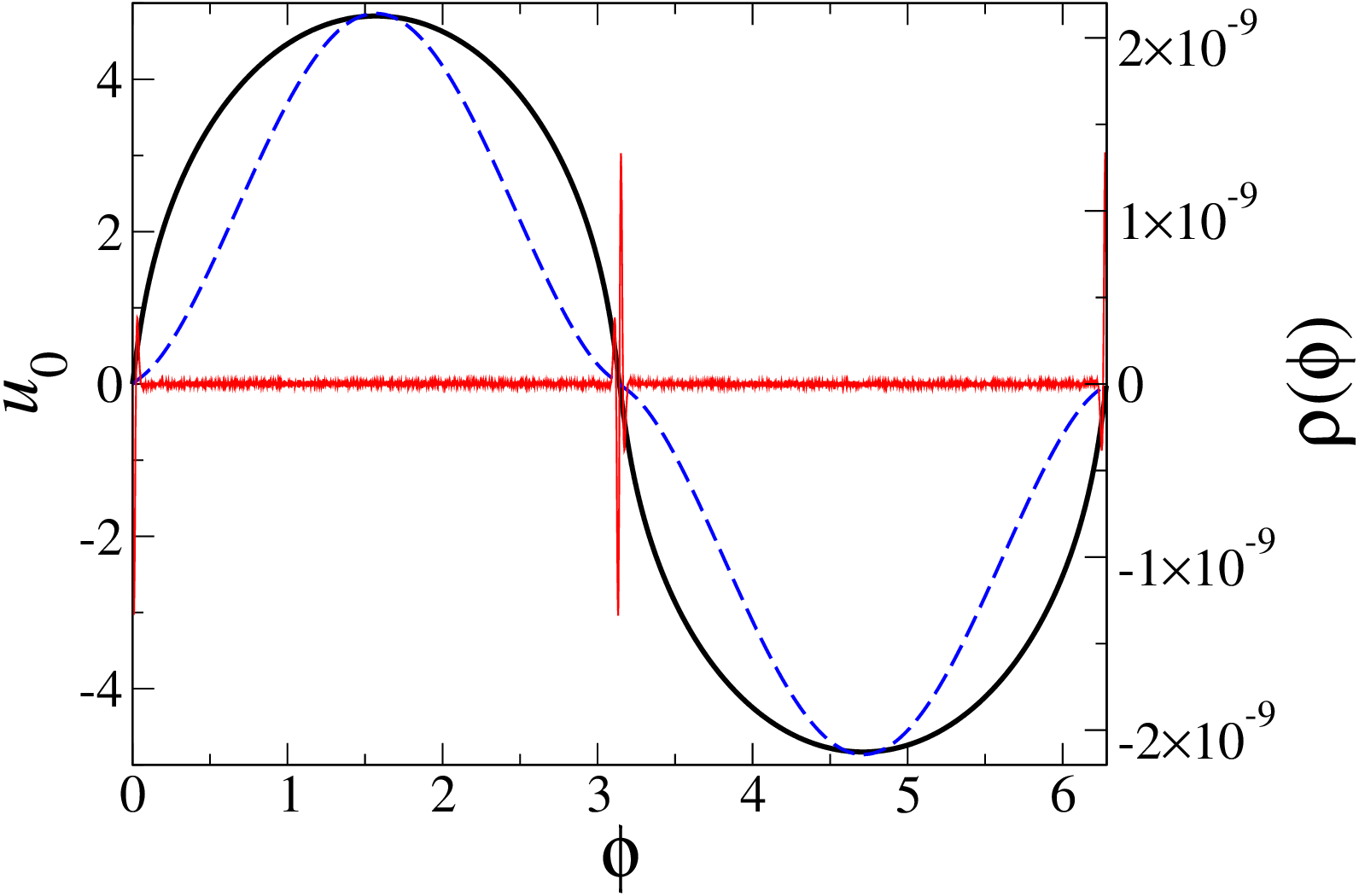}
\caption{Left vertical axis:
Stationary solution $u_0$ (thick full line)
of Eq.~\eqref{nonc_alpha_eq} with $\alpha=2$ and $\lambda = 102.4$, corresponding
to $q =0.061359$. The parameters of the  IFABM4 algorithm are $N=4096$, $\Delta t=0.1$, $\delta t = 0.1$,
$\Delta\rho_{min}=10^{-4}$, and a stopping condition $\rho_\infty < 5\times 10^{-9}$.
Right vertical axis: residue $\rho(\phi)$ before (blue, thin line)
and after (red, dashed line) the
Newton-Raphson minimization.}
\label{fig_Newton-Raphson}
\end{center}
\end{figure*}

The IFABM4 method  has been used in this paper to study the time evolution of Eq.~\eqref{nle} 
and to characterize the coarsening exponent by measuring the mean wavelength of the emerging pattern, $L(t)$,
as a function of time $t$. In order to obtain a reliable  estimate of $n$,
the function  $L(t)$ must be averaged over several  realizations, each one
with a different initial condition. 

Stationary solutions of nonlinear PDEs are normally found with specialized approaches that 
have better performances
compared to any  numerical integration  of the dynamics of Eq.~\eqref{nle}. 
Although   point collocation or multigrid methods~\cite{numerical_recipes} are less computationally expensive than
our IFABM4 code, we decided to rely on the same algorithm to highlight the advantages  of a strategy  based on the
the phase-diffusion equation.
Moreover, even if it is well established that  the integrating factor technique
can lead to a wrong estimation of the fixed points of nonlinear ODEs \cite{Cox:2001} and more accurate
algorithms have been already developed for ODEs \cite{Cox:2001} and PDEs \cite{Kassam:2005}, such as Eq.~\eqref{nle},
a small modification of the IFABM4 algorithm  allowed us to find the stationary solutions $u_0$
with a reasonable precision (enough for our purposes).

Starting from an initial guess function $u(x,0)$, the dynamics of Eq.~\eqref{nleFourier} leads to
the closest stationary solution. Obviously this can happen only when at least one $u_0$ exists for
the fixed value $\lambda$ and for the chosen parameters ($\alpha,\tau,\dots$)
considered during the integration  of Eq.~\eqref{nleFourier}.
The error committed in the estimation of the  stationary solution is easily quantified by inspecting the
 residues
\begin{equation}
\begin{array}{lcl}
\rho(\phi) &=& \mathcal{F}^{-1}\left[\omega_k u_{k} \right]_\phi +{\rm N}\left[ u(\phi)\right], \\[5pt]
\rho_\infty & = & \max \left|\rho(\phi) \right|, \\[5pt]
\rho_2 & = & \int_0^{2\pi} d\phi \rho^2 ,
\end{array}
\end{equation}
during the numerical integration of the PDE.
As shown in Fig.~\ref{fig_timestep}(a), for a fixed value of $\Delta t$ we observe that
$\rho_\infty$ saturates at an asymptotic value depending on
the magnitude of the time steps.
Even if starting the numerical integration of Eq.~\eqref{nleFourier} with a tiny $\Delta t$
  ensures  a small asymptotic residue, this will increase enormously the computational cost
required to find $u_0$,  therefore impairing  the applicability of this method to many practical cases.

\begin{figure*}[t!]
\begin{minipage}{0.495\textwidth}
\epsfig{file=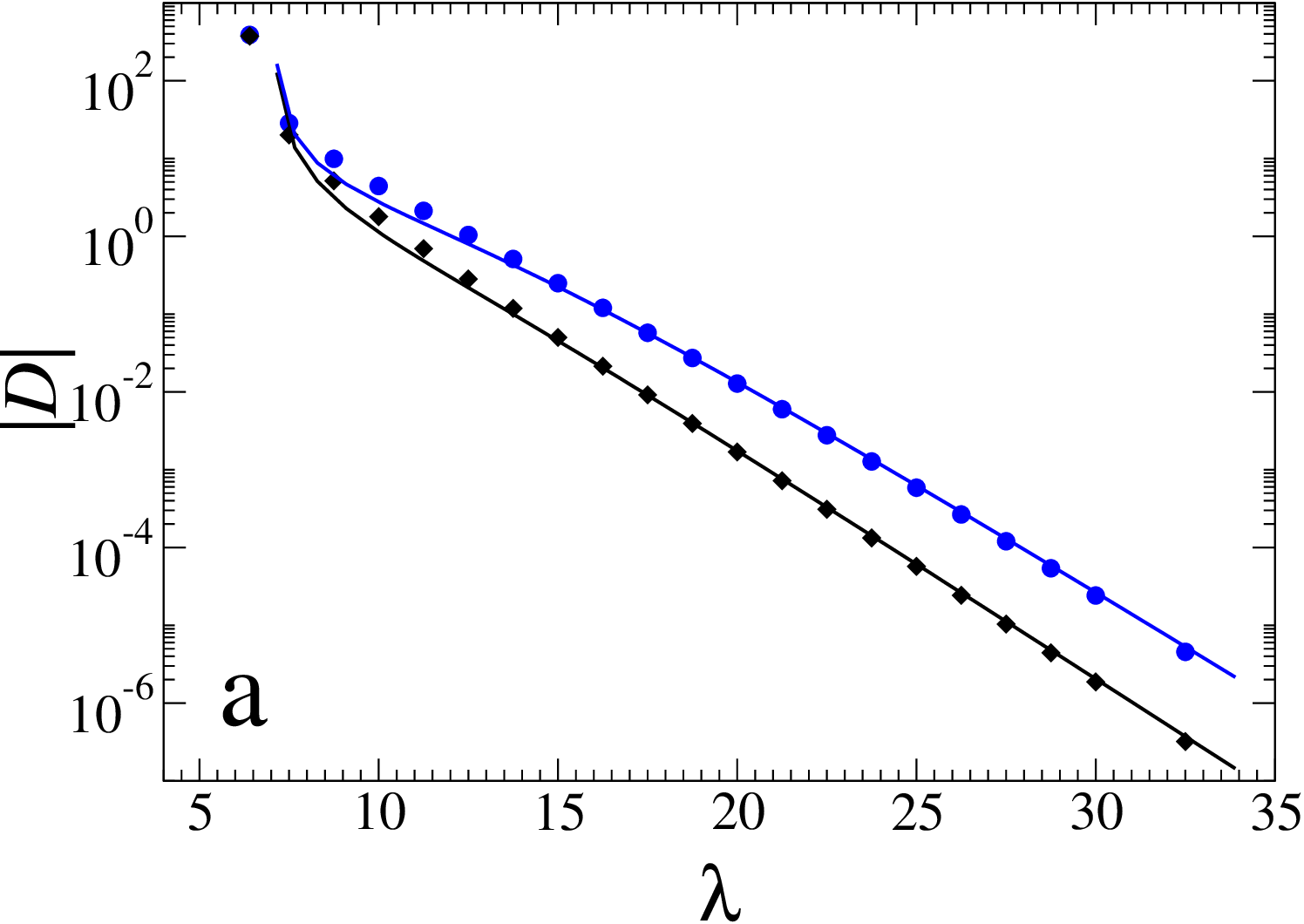,height=6cm,clip=} 
\end{minipage}
\begin{minipage}{0.495\textwidth}
\epsfig{file=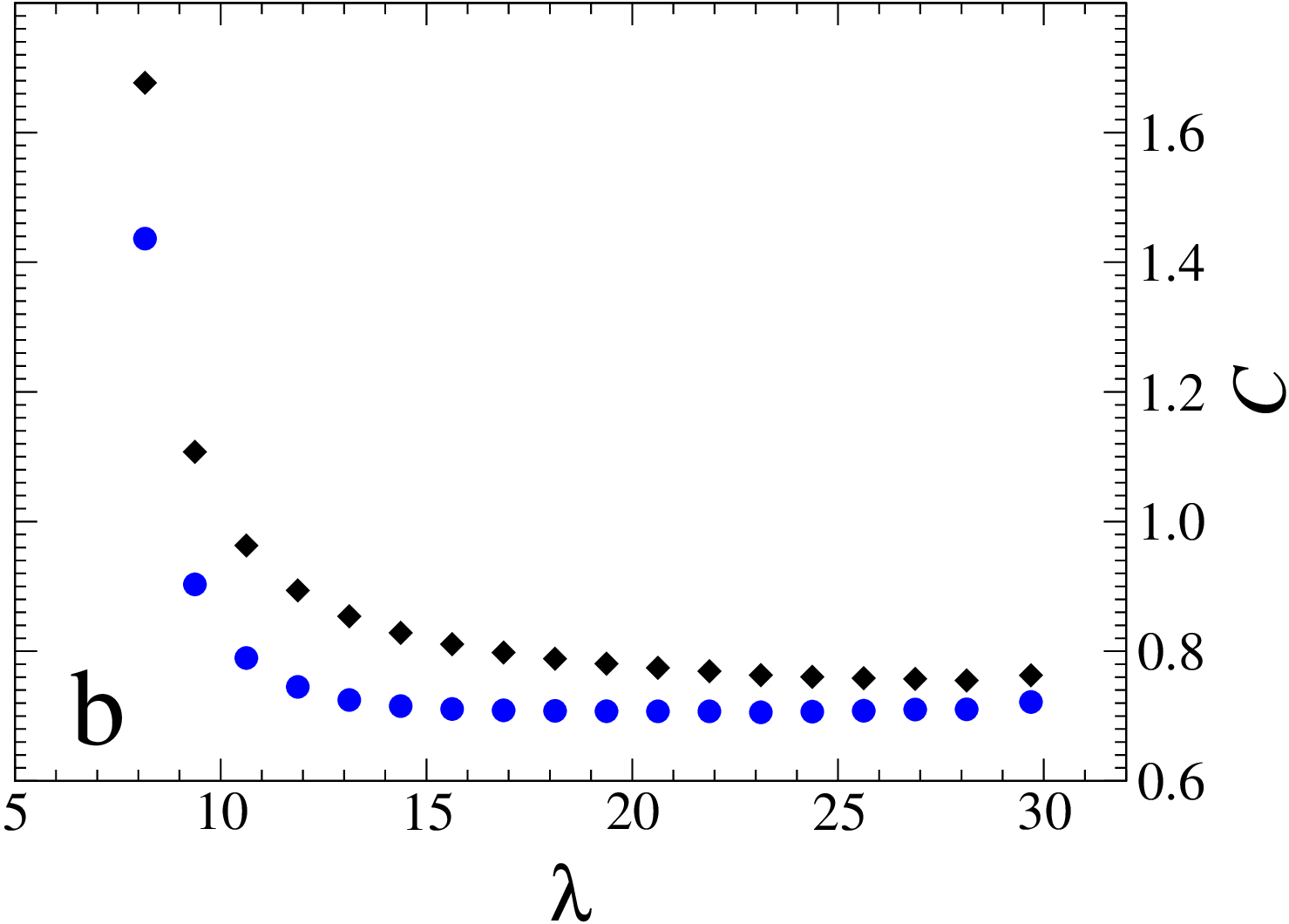,height=6cm,clip=}
\end{minipage}
\caption{ \label{fig_coarsening1}
(a) Phase diffusion coefficient $D$ estimated  with reduced expressions \eqref{Dred_GL}
and \eqref{Dred_CH} for the Ginzburg-Landau equation \eqref{GL_eq} (blue circles)
and for the Cahn-Hilliard equation \eqref{CH_eq} (black rhombi), respectively. 
Solid lines represent values of $D$ calculated from Eqs. \eqref{GL_A}  and \eqref{CH_A}.  
Note that for these two cases $D$ is negative until the sensibility of the algorithm (in this case
of order $10^{-7}$). The stationary solutions $u_0$
have been obtained with a mesh size $\Delta x = 0.025$,
a varying time step $\Delta t$ decreasing from $1$ to $10^{-6}$, and a maximum residue
$\rho_m = 10^{-8}$. The $q$ derivatives in \eqref{Dred_GL} and \eqref{Dred_CH}
are estimated by evaluating   four equally spaced  $u_0$ around $q = 2\pi/\lambda$
with a $\Delta q = 2.5 \times 10^{-3}$.
(b) Estimation of the coefficient $C$ [$\lambda = C^{-1}\ln(t)$]
 for the Ginzburg-Landau equation
\eqref{GL_eq} (blue circles) and the Cahn-Hilliard equation \eqref{CH_eq} (black rhombi).
This coefficient is estimated through a linear regression of $\ln\big(|D|/\lambda^2\big)$
by applying a sliding window binning four consecutive values of $|D|$ from the left figure.
The wavelength   $\lambda$ is  the average of  the different  wavelengths
of the points  binned together.
All units are arbitrary.}
\end{figure*}

A possibility to overcome this trade-off between accuracy and computational effort arises when we take into
account that here our goal is not to follow the real dynamics of Eq.~\eqref{nleFourier} but rather to obtain  a reliable
estimate of its stationary solutions. The numerical convergence of $u(t)$ to $u_0$  can be   increased
by reducing dynamically the value of $\Delta t$. Actually, by monitoring the residue and its variation
$\Delta \rho^m_\infty = |\rho^{m}_\infty - \rho^{m-m_c}_\infty|/\rho^{m}_\infty$ (here the superscript $m$ stands for
the discrete time)
every predefined number of iteration steps $m_c$, we are able to sense when the dynamics
of the PDE has reached the stationary state for a given $\Delta t$.
In fact, a drop of $\Delta \rho^m_\infty$ below a given threshold $\Delta \rho_{min}$ means that the
residual error cannot be diminished anymore, signaling that  is the appropriate moment to reduce $\Delta t$
 by multiplying its value for a factor $\delta_t < 1$. After every reduction of the time step,
the predictor-corrector scheme has to be warmed up by a method of equal accuracy, such as a fourth-order 
Runge-Kutta (see Appendix~\ref{ax_pseudo} for its numerical implementation).
In general, the values of $\Delta \rho_{min}$ and $\delta_t$ have to be inferred  empirically  for
each problem: for our analysis, we have used $\Delta \rho_{min} = 10^{-4}$ and $\delta_t = 0.1$.
Additionally, $m_c$  must be a large number in order
to limit the computational overhead required for the estimation of the residue and its variation.
For one-dimensional problems a value of  $m_c$ between $10^3$ and $10^4$ is a reasonable choice.
As an example, in Fig.~\ref{fig_timestep}(b) we report a comparison between the constant and the variable time stepping
in case of the  Cahn-Hilliard equation \eqref{CH_eq}.

In some cases the solution found by the pseudospectral algorithm is not accurate enough to ensure
a good  estimation of the coarsening law, through the evaluation
of $D(q)$. This usually happens for large system sizes,
so that the algorithm converges slowly,
or when the variations of the  stationary solution are extremely  localized and the resolution of the
numerical method is insufficient to resolve these jumps  (spectral methods are not
suitable to handle discontinuities  like shocks).
To improve the residue of  $u_0$ we have employed a Newton-Raphson (NR) method applied to the
finite difference discretization (in $\phi$ space) of the equation $\mathcal{N}[u] = 0$.
To be more precise, in case of an operator $\mathcal{N}$
that can be written as a second-order derivative  of another nonlinear operator $\mathcal{N}_r$
(that is $\mathcal{N}[u] = -\partial_x^2 \mathcal{N}_r[u]$)
we have found the solutions of equation $\mathcal{N}_r[u] = 0$ and then subtracted
to these functions   their mean value $\langle u_0 \rangle$.
In fact, these type of nonlinear operators are conservative so that
the mean value of $u$ is preserved by the  dynamics of the equation, and, additionally,
we can deal with an operator that is less stiff than the original one.
Besides, it is sufficient to find the
stationary solution of the GL equation \eqref{GL_eq} to compute the phase diffusion coefficient of the CH equation
\eqref{CH_eq} (the same  for the $\alpha$ models).
In the following, as an example,  we detail how we have implemented the Newton-Raphson method to find the
stationary solutions of the conserved Kuramoto-Sivashinsky equation.

The dynamic evolution of the cKS equation presents several numerical difficulties,
which can be ascribed to the nature of its stationary solutions:
a sequence of arcs of parabola whose amplitude grows quadratically with $\lambda$, while
their joining regions have a diverging curvature and a vanishing size~\cite{CKS,Csahok:2000}.
This leads to the blow up of the integration scheme for large time
steps. More specifically, the numerical integration of a
 system with $\Delta x = 0.1$ and $N = 2048$, i.e., $L = 204.8$,
must be performed with our IFABM4 code by  choosing  $\Delta t = 10^{-5}$ in order to prevent
the blow up at $t_{max}\sim 10^3$.
For this reason, the phase diffusion method is very suitable to estimate numerically the coarsening law
for the cKS equation. In this case, instead of performing  a  first estimation of the stationary solutions $u_0$
by means of the pseudospectral algorithm, we have solved directly the  equation
\begin{equation}
\label{cKS_phi_eq}
\left(1-\tau q \partial_\phi + q^2  \partial_\phi ^2 \right) u - q^2 \left(  \partial_\phi u\right)^2 = 0,
\end{equation}
by discretizing the differential operators with sixth-order centered finite differences. In this way
Eq.~\eqref{cKS_phi_eq} is transformed in a set of $N$ nonlinear equations in the space of the
variables $u_i$  (with $i=1,\dots,N$) that is solved by finding the zeros of the set of functions
\begin{equation}
F_i = u_i +q^2 B_i - q A_i\left(\tau +q  A_i \right),
\end{equation}
where $A_i, B_i$ are given by
\begin{align}
\label{A_i}
\hskip -25pt A_i = \left(\partial_\phi u\right)_i &= \dfrac{1}{\Delta\phi}
 	\,	\sum_{j=1}^3 a_j \left(u_{i+j} - u_{i-j} \right), \\ 
\label{B_i}
\hskip -25pt  B_i = \left(\partial_\phi^2 u\right)_i &= \dfrac{1}{\Delta\phi^2}
   \,\left[ b_0 u_i+  \sum_{j=1}^3 b_j \left(u_{i+j} + u_{i-j} \right)\right],
\end{align}
and the coefficients $a_j,b_j$
are listed in Table  \ref{tab_coeff}. These vectors are built  by taking into account the periodic
boundary conditions of $u_i$.
The Jacobian $J$ of this system of equation is needed   in order to solve the problem. The elements
of this matrix are $J_{i,j} = \partial_{u_j} F_i $, which is  a band matrix with
\begin{equation}
\begin{array}{lcl}
J_{i,i} &=& 1+\dfrac{q^2 b_0}{(\Delta\phi)^2}, \\[15pt]
J_{i,i\pm j} &=& (\Delta\phi)^{-2}\, \left[q^2 b_j \mp q a_j\left(\tau\Delta\phi + 2q A_i \right)\right],
\end{array}
\end{equation}
where $j=1,2,3$. Through  the Matlab\textsuperscript{\textregistered} function \verb+fsolve+,
the values of $u_i$ for which  $F_i = 0$ are readily  found by means of
the standard large-scale method implemented in the package, i.e.,
the \verb+trust-region-reflective+ algorithm (for details see Refs.~\cite{Coleman:1994,Coleman:1996}
and the Matlab\textsuperscript{\textregistered} documentation).
Some  stationary solutions $u_0$ of the cKS equation for $\tau=0$
are reported in Fig.~\ref{fig_u0_v}(a).

\begin{figure*}[t!]
\begin{minipage}{0.495\textwidth}
\epsfig{file=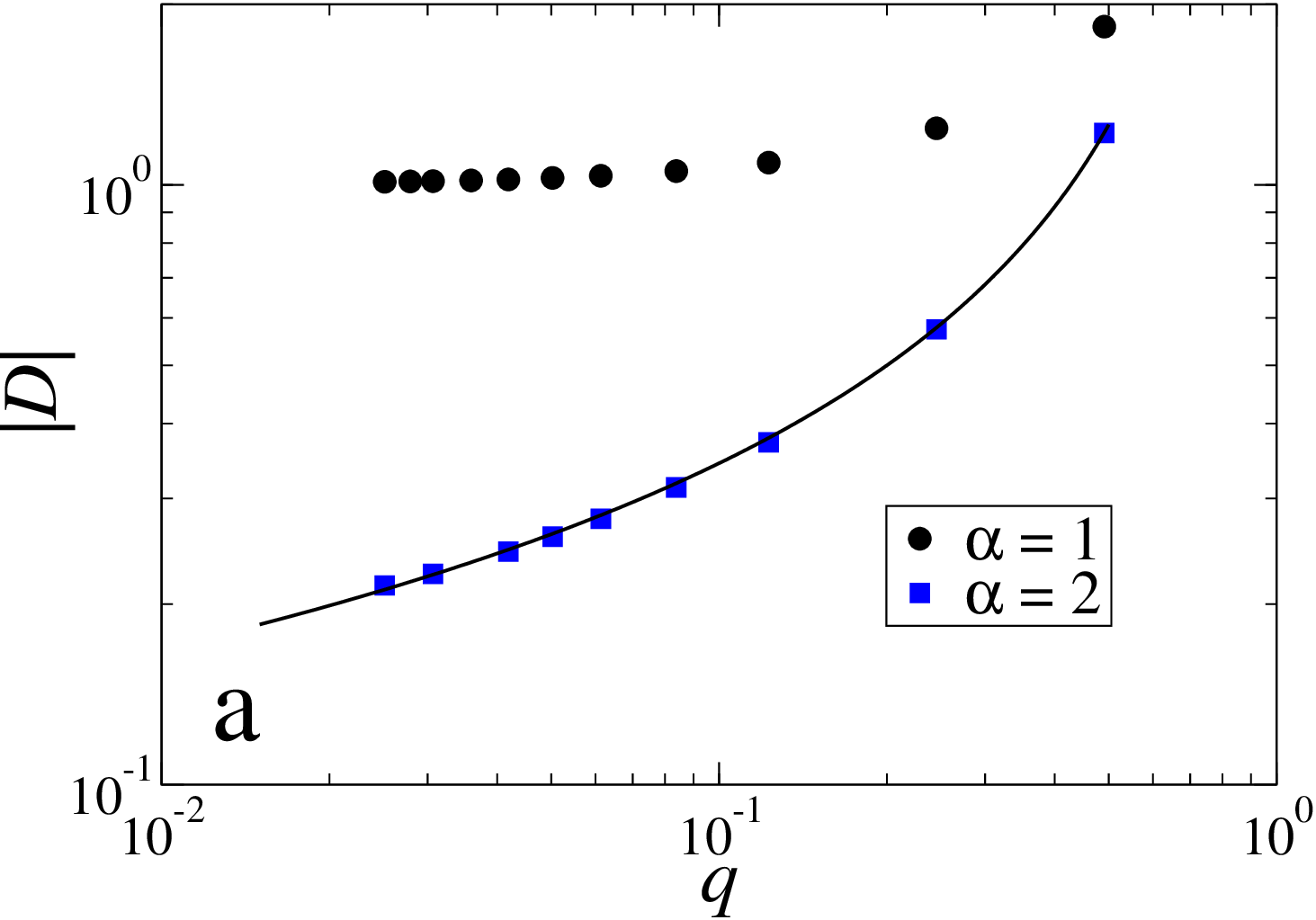,height = 6cm,clip=} 
\end{minipage}
\begin{minipage}{0.495\textwidth}
\epsfig{file=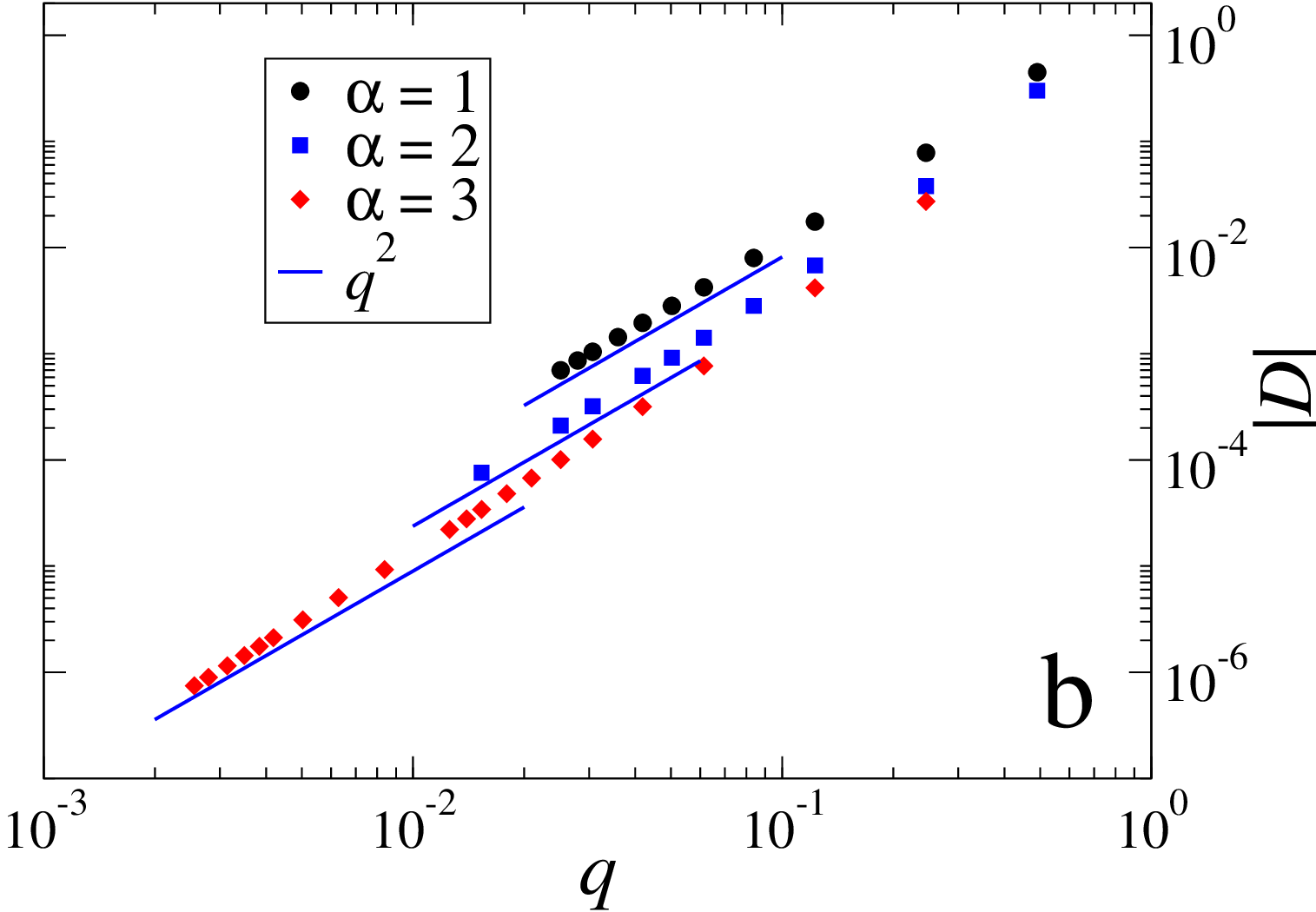,height = 6cm,clip=}
\end{minipage}
\caption{ \label{fig_coarsening2}
(a) Modulus  of the phase diffusion coefficient of the non-conserved $\alpha$ models  \eqref{nonc_alpha_eq}
for different  values of the parameter $\alpha$. The stationary solutions have been obtained through two steps:
(i) a first approximation of $u_0$ is obtained
 by employing  the IFABM4 algorithm up to  a  precision of about $\rho_m \sim 10^{-7}$,
(ii) the final value of the stationary solution is found by refining the result of the previous step through
the Newton-Raphson method.
We used a spatial discretization $\Delta \phi = 0.025$ and $\Delta q = 1.0\times 10^{-4}$ for $\alpha = 1,2$,
whereas $\Delta \phi = 0.05$ and two different  $\Delta q = 1.0\times 10^{-4}, 2.5\times 10^{-5}$ ensuing a ratio
$\Delta q/q < 10^{-2}$  for the case $\alpha = 3$ (see right figure).
The coefficient $D$ has been estimated  through Eq.~\eqref{Dred_GL}. According to 
previous  analytical calculations  \cite{Politi:2006}
 its asymptotic value  saturates to a constant value, i.e., $|D|\sim 1$, when $\alpha \leq 2$.
The solid line is a fit $|D|=a_0/(a_1 + \ln(q))$, showing the logarithmic correction for $\alpha=2$.
(b) Estimation of $|D|$ by means of  \eqref{Dred_CH} for the conserved $\alpha$ models \eqref{c_alpha_eq} with $\alpha=1,2,3$.
We used the same stationary solutions
found for the non-conserved $\alpha$ models. The expected power law for these  conserved models is $|D|\sim q^2$  \cite{Politi:2006}.
The blue solid lines are guides to  eyes with slope equal to two.
All units are arbitrary.}
\end{figure*}

To conclude this section we present one case in which the Newton-Raphson method has been used to decrease
the residue of the stationary solution found by our IFABM4 code. As working example we consider
the non-conserved $\alpha$ model  \eqref{nonc_alpha_eq} with $\alpha=2$. In Fig. \ref{fig_Newton-Raphson}
we show the solution $u_0$ (black line)
and the residue $\rho(\phi)$ before (blue, thin line) and after
(red, dashed line) the NR
minimization.
The residues of the starting guess  (blue line) had
\mbox{$\rho_\infty =  2.14\times 10^{-9}$} and \mbox{$\rho_2 =  1.35\times 10^{-10}$}, but,  after the minimization step, these
residues reduced to \mbox{$\rho_\infty = 1.34 \times 10^{-9}$} and $\rho_2 = 1.12\times 10^{-11}$.
Note the drastic reduction of $\rho(\phi)$ in the regions characterized by small values  of $\partial_\phi u_0$
leading to a $\rho_2$ that is one order of magnitude smaller than before the Newton-Raphson minimization.

\begin{table}[!h]
\begin{tabular}{|c|r|r|r|r|r|}
\hline
index $i$ & $0$  & $1$ & $2$  & $3$  & $4$ \\ 
\hline
$a_i$ & - & $3/4$ & $-3/20$ & $1/60$ & - \\ 
\hline 
$b_i$ & $-49/18$  & $3/2$ & $-3/20$ & $1/90$ & -  \\ 
\hline 
$c_i$ & -  & $-61/30$ & $169/120$ & $-3/10$ & $7/240$ \\ 
\hline 
$d_i$ & $91/8$  & $-122/15$ & $169/60$ & $-2/5$ & $7/240$ \\ 
\hline
\end{tabular} 
\caption{Coefficients of the sixth-order finite difference discretization of the differential operators used in
the Newton-Raphson minimization.}
\label{tab_coeff}
\end{table}


\begin{figure*}[t!]
\begin{minipage}{0.495\textwidth}
\epsfig{file=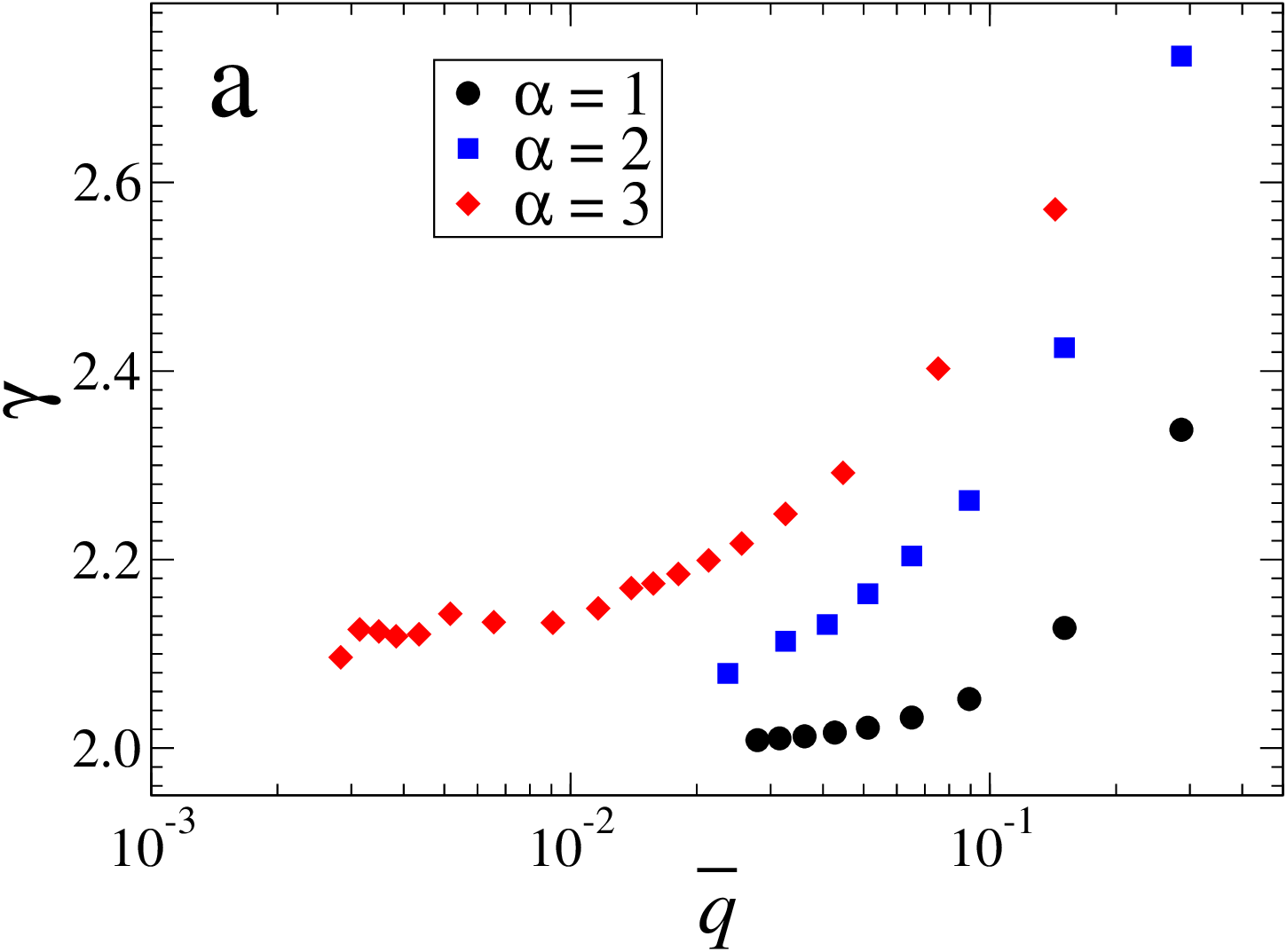,height = 6cm,clip=} 
\end{minipage}
\begin{minipage}{0.495\textwidth}
\epsfig{file=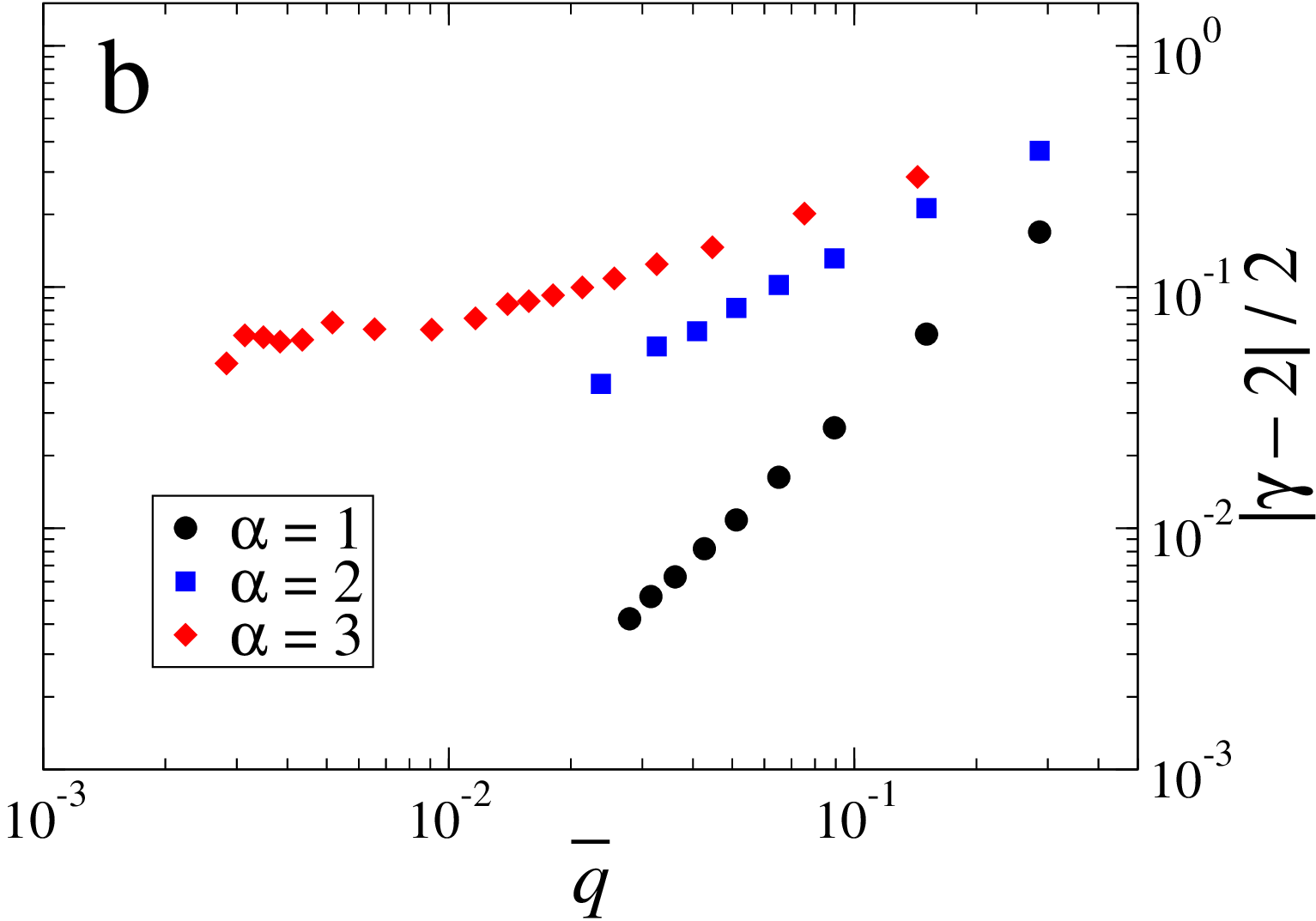,height = 6cm,clip=}
\end{minipage}
\caption{ \label{fig_coarsening3}
(a) Estimation of the power-law behavior of the absolute value of the phase diffusion coefficient
of  the conserved $\alpha$ models for different values of $\alpha$. The exponent $\gamma$ (that is $|D| \sim q^\gamma$) is
computed by a linear regression   of three consecutive points of each curve of
Fig.~\ref{fig_coarsening2}(b).
The center of the sliding window $\bar{q}$ is the mean value of the abscissa of these points.
(b) Relative error of the estimated exponent $\gamma$
as function of the mean value of the window position $\bar{q}$. Note the different convergence rate for every $\alpha$.
All units are arbitrary.}
\end{figure*}

\subsection{Solution of the adjoint operator and calculation of $D$}

The solution of the adjoint problem $\mathcal{L}_0^\dagger[v]=0$
is easily found through the   finite difference discretization introduced in the previous
section. Again, let us consider the conserved Kuramoto-Sivashinsky equation.
Taking into account that in  Eq.~\eqref{eq_adjoint} we  have derivatives up to  fourth order,
first and second derivatives  are discretized according to Eqs.~\eqref{A_i} and \eqref{B_i}, and
third and fourth derivatives are given by
\begin{align}
&\hskip -25pt \left(\partial_\phi^3 v\right)_i = (\Delta \phi)^{-3}\, \sum_{j=1}^{4} c_j\left( v_{i+j} - v_{i-j}\right), \\[5pt]
&\hskip -25pt  \left(\partial_\phi^4 v\right)_i = (\Delta \phi)^{-4}\, \left[d_0 v_i
		+  \sum_{j=1}^{4} d_j\left( v_{i+j} + v_{i-j}\right)\right],
\end{align}
where the coefficients $c_j$ and $d_j$ are listed in Table~\ref{tab_coeff}.

The derivatives  $\partial_\phi^2 u_0$ and $\partial_\phi u_0$ are computed in Fourier space.
The elements of these vectors multiply the constant coefficient of the
finite difference discretization of
the  differential operators of Eq.~\eqref{eq_adjoint}, thus forming
 the sparse matrix $M$, which represent the discrete  version of Eq.~\eqref{eq_adjoint}.
Then, by means of   the Matlab\textsuperscript{\textregistered} function
\verb+eigs+ (that  is based on  the Fortran library \verb+ARPACK+), we search for
the eigenvalues of $M$,  which are in modulus
closer to zero and
their associated eigenvectors.
The eigenvector $\hat{v}$ related to  the smallest eigenvalue is finally  normalized in order to have
a solution $v_0$ with zero mean and amplitude equal to one
\begin{equation}
v_0 = \dfrac{\hat{v} - \langle \hat{v}\rangle}{\max(|\hat{v}|)}.
\end{equation}
Some functions  $v_0$ for the cKS equation for $\tau=0$
 are reported in Fig.~\ref{fig_u0_v}(b).

The values of the phase diffusion coefficient $D$ are  computed  according to
 Eqs.~\eqref{Dred_GL}, \eqref{Dred_CH}, and
\eqref{eq_D}.  As for the solution of the adjoint operator, in these equations
 $\phi$ derivatives are calculated  through Fourier differentiation, whereas
$q$ derivatives are discretized by a fourth-order centered finite differences, i.e.,
\begin{equation}
\begin{split}
\partial_q u_0(q) = \dfrac{1}{12\Delta q}
	\Big[& -u_0(q+2\Delta q)+8 u_0(q+\Delta q) \\
		&\hskip 5pt - 8 u_0(q-\Delta q) + u_0(q-2\Delta q)  \Big].
\end{split}
\end{equation}
These four additional stationary solutions, which are equally distributed around $u_0(\phi;q)$ and
separated by a factor $\Delta q$,
are found by keeping constant the
number of discretization points but changing the length of the system $\lambda$, that is  $2\pi/q$. In this
manner we are able to directly compare the  values of these four functions and
 readily obtain the discrete representation of  $\partial_q u_0(\phi;q)$.
The integrals involved in the calculation of $D$ are evaluated   numerically by the extended
Simpson's rule \cite{numerical_recipes}.

The functional relation between  the phase diffusion coefficient and the wavelength
is found by computing $D$ at discrete increasing  system size (keeping constant
the spatial discretization in $\phi$ space).
After the estimation of $D$ for a given $q_1$, the  stationary solution for $q_2 < q_1$
is computed  by using the information provided by $u_0(q_1)$. The starting guess we used to
determine $u_0(q_2)$ was a vector of $N_2=2\pi/q_2\Delta \phi$ Fourier modes formed by
the Fourier transform  of $u_0(q_1)$ for the first $N_1=2\pi/q_1\Delta \phi$ elements and zero
padded for the others $N_2-N_1$. Moreover, the amplitude of this initial guess has been
adjusted according to the previously observed solutions.
In fact, during  the estimation of $u_0$ for decreasing values of $q$, we record  the amplitude of these
solutions and we are able to forecast the amplitude of the successive stationary solution
(at least roughly). In this manner  we are able to reduce the computational cost needed to
estimate the new value of $D$.
Finally, after the estimation of $D$ for a set of increasing wavelength,
the coarsening law follows by the inversion of   relation \eqref{phase_diffusion},
which leads to   $L(t)\sim \sqrt{|D(L)| t}$.

\begin{figure*}[t!]
\begin{center}
\begin{minipage}{0.495\textwidth}
\includegraphics[height = 6cm,clip=]{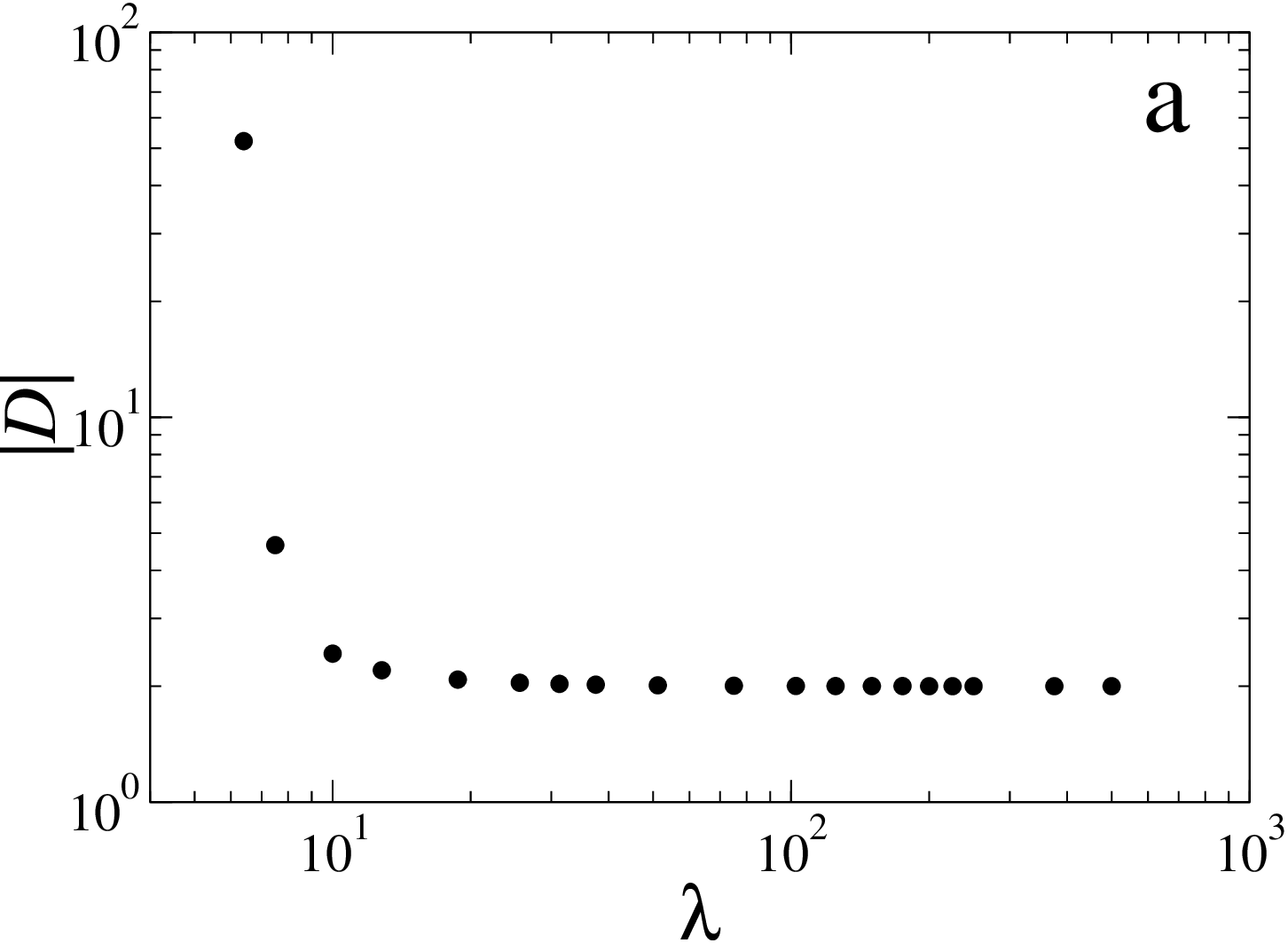} 
\end{minipage}
\begin{minipage}{0.495\textwidth}
\includegraphics[height = 6cm,clip=]{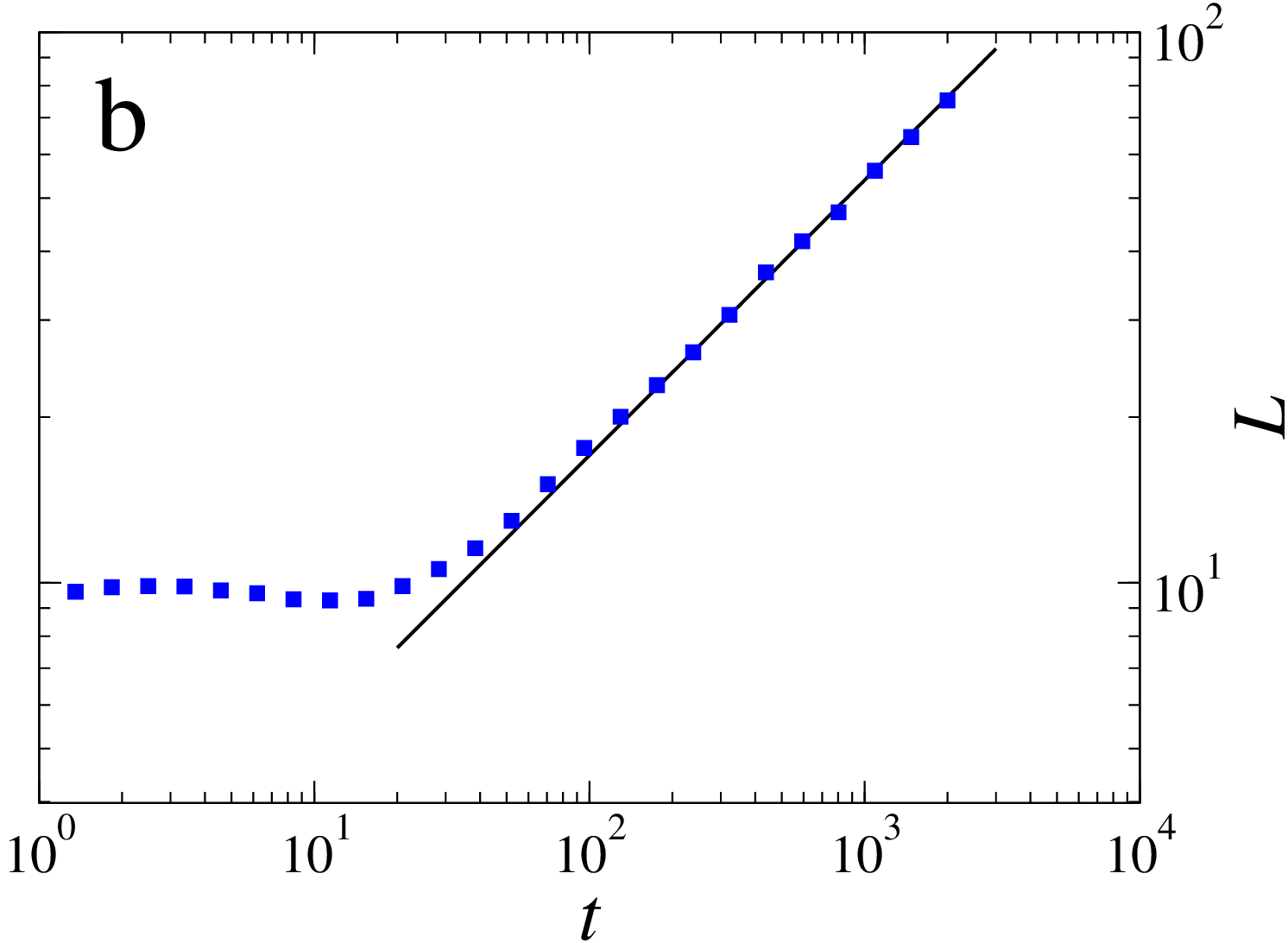}
\end{minipage}
\caption{(a) Absolute value of the phase diffusion coefficient $D$ of the cKS equation \eqref{cKS_eq}
in the case of $\tau = 0$
for different values of $\lambda$. We have used a fixed mesh discretization  $\Delta \phi = 0.025$
and a variable $q$ discretization $\Delta q = 1.0\times 10^{-3},2.5\times 10^{-4}, 1.0\times 10^{-4}$ ensuing a ratio
$\Delta q/q < 10^{-2}$ for the different values of $\lambda$.
(b) Dynamical evolution of Eq.~\eqref{cKS_eq} for  $\tau = 0$ with simulation
parameters $N=2048$, $\Delta \phi = 0.1$, and $\Delta t = 10^{-5}$. The values of
$\lambda$ are averaged over $50$ different random initial  conditions  whose
elements are sampled  from a Gaussian distribution with zero mean and unit variance.
The black solid line is a guide to eyes with slope $1/2$. All units are arbitrary.}
\label{fig_D_dyn_cKS}
\end{center}
\end{figure*}

\section{Results}
\label{sec_results}

We now report and discuss the results we obtained using
the numerical methods described in previous sections.
Let us start with the simplest and well known models: the
GL model (\ref{GL_eq}) and its conserved version, the
CH model (\ref{CH_eq}). They admit analytical solutions based on the Jacobi elliptic function 
${\rm Sn}(x;p)$~\cite{Villain_Guillot:08},
allowing us to test our numerical procedure with a controlled  result. This family of solution
is parametrized by the elliptic modulus  $p \in [0,1]$ (for more details see Appendix~\ref{ax_GL}) 
and take the form
\begin{equation}
u_0(x) = A \,  {\rm Sn}\left(\dfrac{4 K}{\lambda} x ; p\right).
\end{equation}
The dependence of the complete elliptic integral of the first kind $K$
and the amplitude $A$ on the elliptic modulus $p$ is reported 
 in Appendix~\ref{ax_GL}, Eqs.~(17) and (18), respectively.
It has  already been shown that  
for these two models  the value of  the phase-diffusion coefficient can be expressed  
as a function of $\lambda$, the amplitude of $u_0$ and the integrals 
$J = \int_0^\lambda  (u_0')^2 \, dx$  for the GL model 
or  $I = \int_0^\lambda u_0^2 \, dx$ for its conserved counterpart~\cite{Politi:2006,PRL} 
\begin{eqnarray}
\label{GL_A}
D = - \dfrac{\lambda^2 ( A - A^3)}{J\, \partial_A \lambda} & \quad {\rm GL\ equation}, & \\
\label{CH_A}
D = - \dfrac{\lambda^2 ( A - A^3)}{I\, \partial_A \lambda} & \quad {\rm CH\ equation}. &
\end{eqnarray}
As depicted in Appendix~\ref{ax_GL},
we can compute all these functions and their derivatives with respect to $A$ through
complete elliptic integrals of first or second kind and Jacobi elliptic functions. Finally, the integrals $I$ and $J$ are
easily evaluated by means of adaptive quadrature so that the values of $D$ are obtained without solving any differential 
equation.

\begin{figure*}[t!]
\begin{center}
\begin{minipage}{0.495\textwidth}
\includegraphics[height = 6cm,clip=]{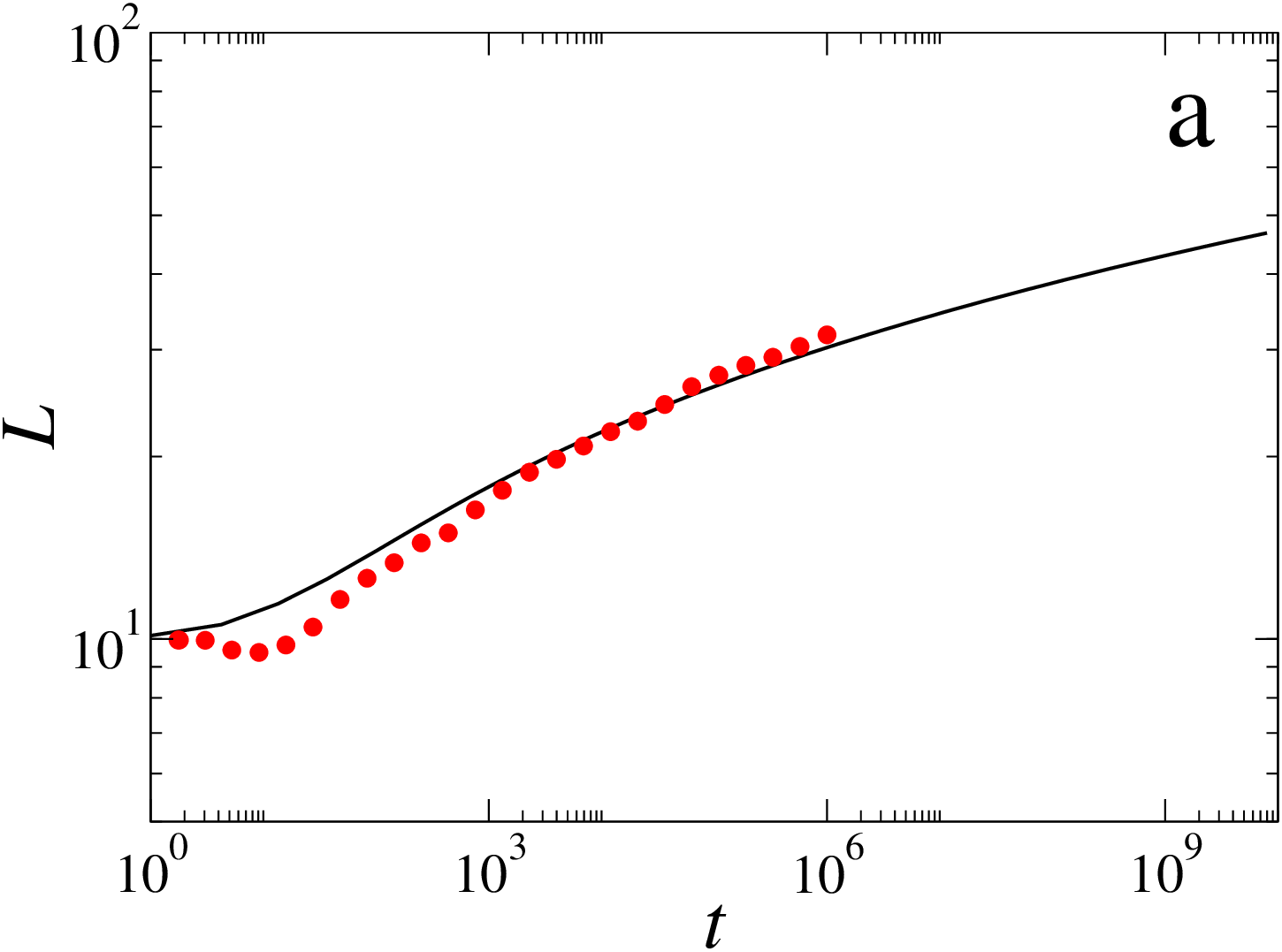} 
\end{minipage}
\begin{minipage}{0.495\textwidth}
\includegraphics[height = 6cm,clip=]{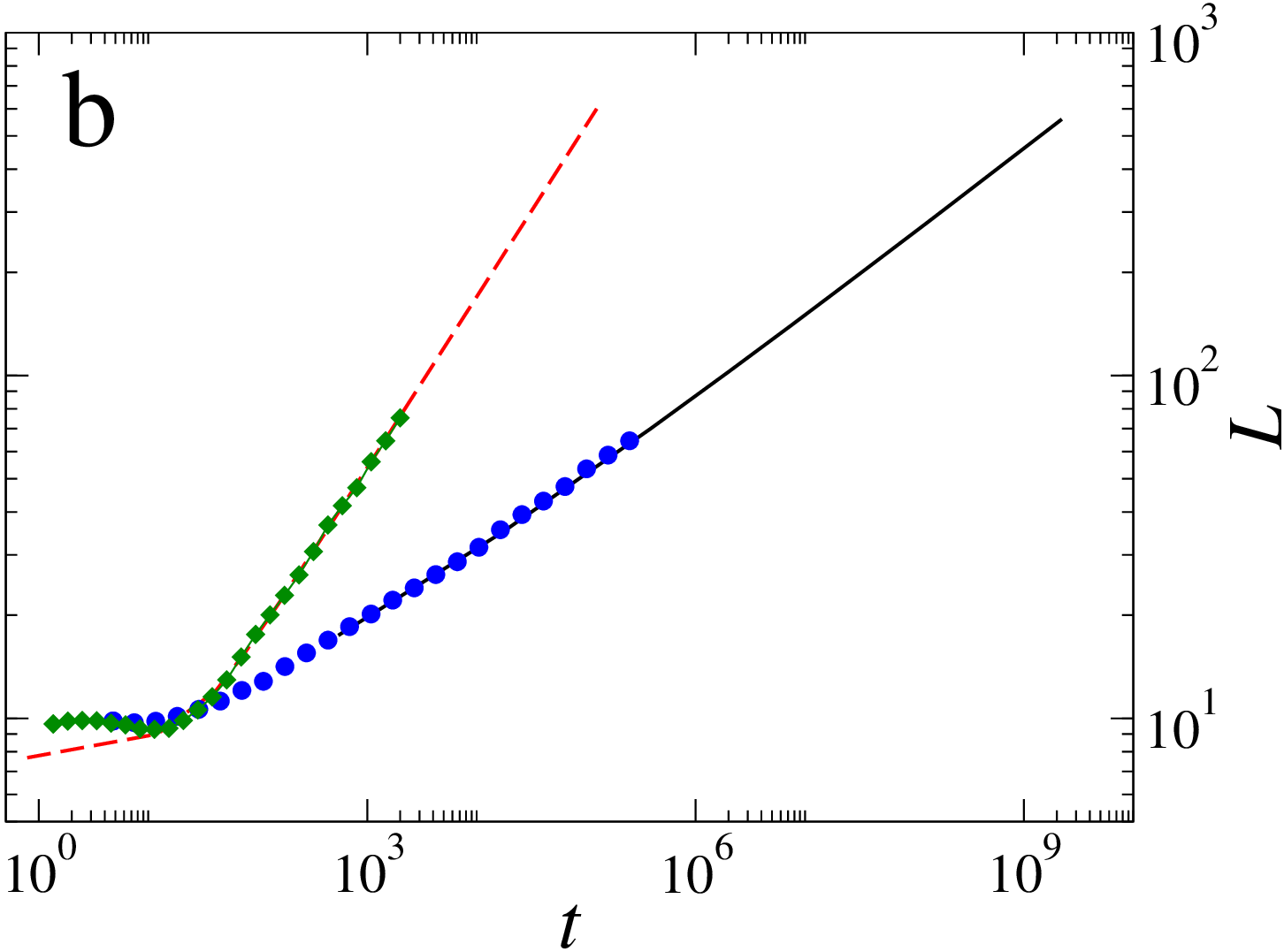}
\end{minipage}
\caption{Comparison between the phase diffusion method and the  numerical integration of the time-dependent
PDE for three different models. (a) Case of logarithmic coarsening,  the Cahn-Hilliard equation. The black line is
 the estimation of $\lambda$ as function of $t$ from the data displayed in Fig.~\ref{fig_coarsening1}(a). 
The red circles are  the numerical integration of Eq.\ \eqref{CH_eq} by means of our  IFABM4 code with
parameters $L = 102.4$, $\Delta x = 0.1$, $\Delta t = 0.05$, and averaged over $50$ different initial conditions
(generated from a normal distribution with zero mean and unit variance). (b) Case of power-law coarsening.        
As for the left panel, lines stand for the estimation of $\lambda$ as function of $t$ from the values of $D(\lambda)$ while
the symbols are obtained from the numerical integration of time dependent equations. 
The black solid line [from data in Fig.~\ref{fig_coarsening2}(b)] and  the blue circles are data for the conserved 
$\alpha$ model Eq.\ \eqref{c_alpha_eq} with $\alpha=2$ and parameters
$L=256$, $\Delta x = 0.25$, $\Delta t = 0.05$, and averaged over $50$ different initial conditions.
The red dashed line and the green rhombi correspond to the conserved Kuramoto-Sivashinsky equation \eqref{cKS_eq}. These
data  are computed (line) and reported (rhombi) from the left and the right panel of Fig.\ \ref{fig_D_dyn_cKS}, respectively.
All units are arbitrary.}
\label{fig_time_comparison}
\end{center}
\end{figure*}

In the absence of noise, the Ginzburg-Landau and the Cahn-Hilliard model display
a logarithmic coarsening, $L(t) \simeq C^{-1} \ln t$, which corresponds,
in the formalism of phase diffusion equation, to a diffusion coefficient
which decreases exponentially with the wavelength of the stationary
solution, $D(\lambda) \simeq e^{-C\lambda}$.
Figure \ref{fig_coarsening1}(a) shows such exponential behavior on a lin-log
scale while Fig.~\ref{fig_coarsening1}(b) shows the convergence of the
constant $C$ to a value of order 0.75 for both models. 
Note that in Fig.~\ref{fig_coarsening1}(a) the  comparison between the values of $D$ estimated numerically
(symbols) and those obtained through Eqs. \eqref{GL_A} and \eqref{CH_A} (lines) 
confirms the validity of our method in a well-controlled situation and its applicability  to other non-trivial cases.  

Next, we consider the nonconserved and conserved $\alpha$ models,
given by Eqs.~(\ref{nonc_alpha_eq}) and (\ref{c_alpha_eq}), respectively.
According to the analytic treatment of the phase diffusion
equation \cite{Politi:2006} and to numerics (taking care of
dangerous caveats \cite{alpha_sim}), we expect that the
coarsening exponent for the nonconserved models is
$n={1/ 2}$ for $1\le\alpha < 2$ and $n=\alpha/(3\alpha -2)$ for
$\alpha >2$. Case $\alpha=2$ has a logarithmic correction,
$L(t)\simeq \sqrt{t}/\ln t$. Conserved models, instead show a
constant coarsening exponent, $n={1/ 4}$, for all $\alpha$.
Figure \ref{fig_coarsening2}(a) considers cases $\alpha=1,2$ for
nonconserved models, therefore we expect a constant $D$ for
$\alpha=1$ and $D\approx 1/\ln q$ for $\alpha=2$. This is exactly what
our numerics shows.
In Fig.~\ref{fig_coarsening2}(b) we consider $\alpha=1,2,3$ for the
conserved models and, as expected, $|D| \simeq q^2$ for small $q$.
The conserved case is analyzed in more details in Figs.~\ref{fig_coarsening3}(a) and \ref{fig_coarsening3}(b). 
On the left we plot the exponent
$\gamma$ defined by $|D| \simeq q^\gamma$, and on the right
we specialize on its convergence to the asymptotic value
$\gamma_\infty=2$. According to the results displayed in  Fig.~\ref{fig_coarsening3}(b), 
for a larger exponent $\alpha$ we observe a slower convergence to the asymptotic $\gamma_\infty$. 
This effect is more pronounced for the case $\alpha=3$ where a crossover region 
between two power laws has been already reported in Ref.~\cite{alpha_sim}. Taking into account that 
this region is of the order $\lambda\sim 10^3$, i.e., $q\sim 0.006$, and for this wavelength 
our algorithm estimates a  $|D|$ smaller than $10^{-5}$, it is difficult to assess a clear 
asymptotic value for $\gamma$ and the relative error $|\gamma - \gamma_\infty|/\gamma_\infty$
saturates to a value around  5\%.

In Figs.~\ref{fig_D_dyn_cKS}(a) and \ref{fig_D_dyn_cKS}(b) we consider the
cKS model (\ref{cKS_eq}). Some numerical and approximated analytical
estimations of the coarsening exponents gave $n=1/2$, i.e.,
a constant phase diffusion coefficient. This is shown on the left,
while the right shows a clear evaluation of the
coarsening exponent through the direct integration of the
dynamical equation. Our results are much cleaner than available results
in the literature.

Finally we compare the computational cost used to estimate the coarsening
exponent by means of the phase diffusion method with the cost of the standard 
method used in the literature, that is, typically,  the direct integration of
the time-dependent PDE. We consider three cases of conserved models  that
are usually stiffer that their non-conserved counterparts due to  high-order spatial
derivatives. Lines in Fig.\ \ref{fig_time_comparison}  are found after the inversion
of the relation $\lambda^2/|D|\sim t$ that leads to the coarsening law $L(t)\sim t^n$. 
In the same figure, symbols  are data obtained from the numerical integration of  
time-dependent PDEs averaged over several realizations of different initial conditions
(see captions for more details). As shown in the figure,
the phase diffusion method allows one to extract the coarsening law for times that 
are  orders of magnitude larger compared to the standard method. Moreover, the times
listed in Table~\ref{tab_time} demonstrate the remarkable performance of  our implementation
of the phase diffusion method. Note that in the three cases we have not employed the same
strategy to find the stationary solutions of the different PDEs. As already explained in the previous sections, for the
Cahn-Hilliard model we have only used our IFABM4 code with variable time stepping,   
for the conserved Kuramoto-Sivashinsky equation we have solved directly Eq.\ \eqref{cKS_phi_eq} with
the Newton-Raphson method, and for the conserved $\alpha$ model we combined these
two methods. We have also verified that the solution of the adjoint problem \eqref{eq_adjoint} for the cKS model 
is much less costly (roughly two orders of magnitude) than the determination of $u_0$.  

\begin{table*}[!t]
\begin{tabular}{|l|c|c|c|c|}
\hline
& Phase diffusion  & Integration & Integration & Ratio (long time)  \\
& (long time) & (short time) & (long time) & Integration/Phase diffusion\\
\hline
CH &  $8.32 \times 10^2$ & $8.54\times 10^4$ & $\sim 4.21 \times 10^9$  &  $\sim 5 \times 10^6$ \\
c$\alpha$-model ($\alpha$=2) & $1.66\times 10^3$ & $3.95 \times 10^4$  & $\sim 7.41\times 10^9$  & $\sim 4 \times 10^6$   \\ 
cKS & $6.07\times 10^2$  & $1.46 \times 10^6$ & $\sim 1.47\times 10^9$ &   $\sim 2 \times 10^6$ \\ 
\hline
\end{tabular} 
\caption{Actual computational time (in seconds) 
required to measure the coarsening exponent $n$ with the phase diffusion method
(second column) and with the numerical integration of the time-dependent PDE (third column). 
The fourth column is an estimation of the computational cost needed by the standard method to reach the same 
time span of the phase diffusion method for the data reported in Fig.\ \ref{fig_time_comparison}. 
The fifth column is the ratio between the estimated time for the standard method and the cost for the phase diffusion
method. The numerical experiments were carried out in a machine equipped with a 
single-processor Intel\textsuperscript{\textregistered} Core{\texttrademark} 2 Duo with a clock frequency of $2.4$ Ghz.}
\label{tab_time}
\end{table*}


\section{Conclusions}
\label{sec_conc}

In this work we have demonstrated that  the phase diffusion method can be a reliable and fast approach
to find the coarsening law of nonlinear 1D partial differential equations.
Its numerical implementation is straightforward and extends the applicability of the method beyond 
the cases  studied in Ref.~\cite{Politi:2006}.
In fact, all these cases allowed an analytical evaluation of $L(t)$, which is not always
possible (see the cKS equation).

Our algorithm permits  us to evaluate the coarsening exponent with a negligible 
computational cost compared to  standard methods, such as the direct  numerical solution of the 
time-dependent PDE (see Table 2).  
Furthermore, it does not suffer from spurious effects associated with the finite size of the system
or from errors arising in the time discretization 
of the PDE, and the result does not need to be averaged over several initial conditions.  

The main limitation of the method is the finite accuracy in the estimation of $D$, which
should be improved for larger times, when coarsening is slow (logarithmically slow or following
a power law with a small exponent $n$). However, in such cases the direct time integration
would require to attain extremely large times. Our discussion and in particular Table 2 shows that the
phase diffusion method is computationally much faster than time integration.

A separate comment should be made for PDEs whose steady solutions are determined by an ODE
of higher order than two. In these cases~\cite{Peletier}, which include the Swift-Hohenberg equation,
we expect more branches of stationary solutions, $\lambda_i(A)$, which makes our analysis
more complicated. 

Finally, in this paper we confined our discussion to one-dimensional PDEs, but we have recently
shown~\cite{2D} that the phase diffusion method can be successfully applied to coarsening processes
in two dimensions  as well. It will be interesting to extend our numerical approach to the bidimensional case,
especially for studying those equations that  can hardly be attacked analytically.

\section*{Acknowledgements}
M.~N. and P.~P. acknowledge support from the Italian Ministry of Research (PRIN 2007JHLPEZ).
\mbox{M.~N.} also acknowledges partial support by MICINN (Spain) Grant No.\ FIS2009-12964-C05-01.
C.~M. acknowledges financial support from CNES (Centre National d'Etudes Spatiales)  and
thanks the Italian CNR for support from their International Exchange Program.


\appendix

\begin{widetext}
\section{Pseudospectral algorithm}
\label{ax_pseudo}

The solution of the linear part of  Eq.~\eqref{nleFourier} suggests the following change of variable
\begin{equation}
z_k(t) = e^{-\omega_k t} u_k(t),
\end{equation}
and the temporal evolution of this new variable  depends only on the nonlinear operator
\begin{equation}
\label{evolutz}
\partial_t z_k(t) = e^{-\omega_k t}\, {\rm N}\left[u(t)\right]_k.
\end{equation}
This is our starting point to discretize  Eq.~\eqref{evolutz} in the time domain by
using one of the standard algorithms available in the literature.
In order to enhance the stability properties of our method we have chosen a fourth order
predictor-corrector method, the Adams-Bashforth-Moulton (IFABM4) scheme.
The predictor step applied to Eq.~\eqref{evolutz} is
\begin{equation}
\begin{split}
z_k(t+\Delta t) = z_k(t) +\frac{\Delta t}{24}\bigg[& 55 e^{-\omega_k t}\, {\rm N}\left[u(t)\right]_k -
	59 e^{-\omega_k (t-\Delta t)}\, {\rm N}\left[u(t-\Delta t)\right]_k   \\[5pt]
&+	37 e^{-\omega_k (t-2\Delta t)}\, {\rm N}\left[u(t-2\Delta t)\right]_k -
	9 e^{-\omega_k (t-3\Delta t)}\, {\rm N}\left[u(t-3\Delta t)\right]_k \bigg],
\end{split}
\end{equation}
that leads to a prediction for $\bar{u}_k$ at the next time step
\begin{equation}
\bar{u}_k(t+\Delta t) = e^{\omega_k \Delta t} \left[u_k(t) + \frac{\Delta t}{24}\left( 55 {\rm N}_0 -
	59 {\rm N}_1 + 37 {\rm N}_2 - 9 {\rm N}_3 \right) \right],
\label{predictor}
\end{equation}
where the functions ${\rm N}_n = e^{n \omega_k \Delta t}\, {\rm N}\left[u_k(t- n\Delta t)\right]_k$
are computed during the time stepping of the algorithm by successive multiplications of the
 factor $\exp(\omega_k \Delta t)$.
As before, the corrector step is applied to the variable $z_k(t+\Delta t)$ and is computed from
the value of the predictor and the value of $u_k$ at three previous  time steps
\begin{equation}
\begin{split}
z_k(t+\Delta t) = z_k(t) +\frac{\Delta t}{24}\bigg[ &9 e^{-\omega_k (t+\Delta t)} {\rm N}\left[\bar{u}(t+\Delta t)\right]_k +
	19 e^{-\omega_k t} {\rm N}\left[u(t)\right]_k  \\[5pt]
&-	5 e^{-\omega_k (t-\Delta t)} {\rm N}\left[u(t-\Delta t)\right]_k +
	 e^{-\omega_k (t-2\Delta t)} {\rm N}\left[u(t-2\Delta t)\right]_k \bigg],
\end{split}
\end{equation}
so that  the update of $u_k$ reads
\begin{equation}
\label{corrector}
u_k(t+\Delta t) = \frac{3}{8}\Delta t\,  {\rm N}\left[\bar{u}(t+\Delta t)\right]_k +
	e^{\omega_k \Delta t} \left[u_k(t) + \frac{\Delta t}{24}\left( 19 {\rm N}_0 -
	5 {\rm N}_1 +  {\rm N}_2  \right) \right].
\end{equation}

Equations  \eqref{predictor} and \eqref{corrector} depend on four previous
values of the function $u_k$, we have to compute the four initial time steps of the
evolution of \eqref{nleFourier} by an integrator of the same accuracy,
for example a classical fourth order Runge-Kutta method (RK4). For the variable $z_k$
the RK4 method reads
\begin{equation}
\label{rkm}
z_k(t+\Delta t) = z_k(t) +\frac{\Delta t}{6}\left(f_1 + 2 f_2 + 2 f_3 + f_4  \right),
\end{equation}
where the different RK4 steps are obtained through a sequence of new variables. The first point in the
RK4 evolution is easily found by evaluating Eq.~\eqref{evolutz} at time $t$ with the variable $u_k(t)$
\begin{equation}
f_1 = e^{-\omega_k t}\, {\rm N}\left[\mathcal{F}^{-1}\left[ e^{\omega_k t} z_k(t)\right]_\phi \right]_k
=e^{-\omega_k t}\, {\rm N}_0.
\end{equation}
Then we introduce the new variable $u_{1k}$ and we find the second RK4 point $f_2$
\begin{equation}
u_1 = \mathcal{F}^{-1}\left[e^{\omega_k \Delta t/2} \left( u_k(t) + \dfrac{\Delta t}{2}\, {\rm N}_0 \right)\right]_\phi,
\end{equation}
hence $f_2 = \exp[-\omega_k(t+\Delta t/2)] \, {\rm N}\left[u_1\right]_k$.
The same procedure is iterated, so that
\begin{equation}
u_2 = \mathcal{F}^{-1}\left[ e^{\omega_k \Delta t/2} \left( u_k(t) + \dfrac{\Delta t}{2}\,{\rm N}\left[u_1\right]_k\right)\right]_\phi, 
\end{equation}
so that $f_3 = \exp[-\omega_k(t+\Delta t/2)]\, {\rm N}\left[u_2\right]_k$, and
\begin{equation}
u_3 = \mathcal{F}^{-1}\left[e^{\omega_k \Delta t} \left( u_k(t) + \Delta t \, {\rm N}\left[u_2\right]_k\right)\right]_\phi, 
\end{equation}
hence  $f_4 = \exp[-\omega_k(t+\Delta t)]\, {\rm N}\left[u_3\right]_k$.
Finally, the update for $u_k$ reads
\begin{equation}
u_k(t+\Delta t)= e^{\omega_k\Delta t}\left[u_k(t)+\frac{\Delta t}{6}\,{\rm N}_0\right]
+\frac{\Delta t}{6}\left[ 2 e^{\omega_k\Delta t/2}\left( {\rm N}[u_1]_k + {\rm N}[u_2]_k\right)+{\rm N}[u_3]_k\right].
\end{equation}
This last equation is used to warm up the predictor-corrector method with the three initial time steps.


\section{Analytical solutions of the Ginzburg-Landau equation}
\label{ax_GL}

The stationary states of the Ginzburg-Landau equation can be expressed
analytically by means of  the Jacobi elliptic function sine-amplitude 
${\rm Sn}(x;p)$~\cite{Villain_Guillot:08}. This family of solutions
is parametrized by the elliptic modulus  $p \in [0,1]$ and take the form
\begin{equation}
u_0(x) = \sqrt{\dfrac{2 p^2}{p^2+1}}  \,  {\rm Sn}\left(\dfrac{x}{\sqrt{p^2+1}} ; p\right),
\end{equation}
and its derivative reads
\begin{equation}
u'_0(x) = \dfrac{\sqrt{2} p^2}{p^2+1}  \,  {\rm Cn}\left(\dfrac{x}{\sqrt{p^2+1}} ; p\right)
	\, {\rm Dn}\left(\dfrac{x}{\sqrt{p^2+1}} ; p\right),
\end{equation}
where ${\rm Cn}$ and ${\rm Dn}$ are the other two Jacobi elliptic functions~\cite{Wolfram}.
The periodicity of these functions is 
\begin{equation}
\lambda = 4K(p) \sqrt{p^2+1},
\end{equation}
where the complete elliptic integral of the first kind is defined according to
\begin{equation}
\label{eq_Kp}
K(p) = \int_0^1 \dfrac{dt}{\sqrt{(1-t^2)[1-p^2 t^2]}}.
\end{equation}
Taking into account that the function ${\rm Sn}\in[-1,1]$, the 
amplitude of $u_0$ is a function of the elliptic modulus 
\begin{equation}
\label{eq_Ap}
A = \sqrt{\dfrac{2 p^2}{p^2+1}},
\end{equation}
so that we are able to express the periodicity of the stationary solution as a function
of its amplitude
\begin{equation}
\lambda(A) = 4\sqrt{\dfrac{2}{2-A^2}}\, K\left( \dfrac{A}{\sqrt{2-A^2}}\right).
\end{equation}
The variation of $\lambda$ with respect to the amplitude of $u_0$ is
\begin{equation}
\partial_A\lambda =\dfrac{4}{A}\left(\dfrac{2}{2-A^2} \right)^{1/2}
\left[\dfrac{1}{1-A^2} \, E\left(\dfrac{A}{\sqrt{2-A^2}}\right) - K\left(\dfrac{A}{\sqrt{2-A^2}}\right) \right],
\end{equation}
where the complete elliptic integral of the second kind is
\begin{equation}
E(p) = \int_0^1 \, dt \sqrt{\dfrac{1-p^2 t^2}{1-t^2}}.
\end{equation}

\end{widetext}


\begin{thebibliography}{100}

\bibitem{Politi:2006} P. Politi, and C. Misbah,
Nonlinear dynamics in one dimension: A criterion for coarsening and its temporal law,
{\sl Phys. Rev. E} {\bf 73}, 036133 (2006).

\bibitem{note1} Coarsening may also occur when there is a minimal wave vector $q_{min}$ below which
$\omega(0)<0$ as in Ref.~\cite{strange_coarsening}. However, this scenario is out of our interest.

\bibitem{PRL} P. Politi and C. Misbah, When Does Coarsening Occur in the Dynamics of One-Dimensional Fronts?,
{\sl Phys. Rev. Lett.} {\bf 92}, 090601 (2004).

\bibitem{note2} A negative $D$ might imply an anticoarsening process as well, in principle. However, here we don't consider this scenario.

\bibitem{Bray}
A.~J. Bray, Theory of phase-ordering kinetics, {\sl Adv. Phys.} \textbf{43}, 357 (1994).

\bibitem{PPJV}
P. Politi and J. Villain, Ehrlich-Schwoebel instability in molecular-beam epitaxy: A minimal model, 
{\sl Phys. Rev. B} {\bf 54}, 5114 (1996).

\bibitem{ATPP}
P. Politi and A. Torcini, Coarsening in surface growth models without slope selection, 
{\sl J. Phys. A: Math. Gen.} {\bf 33}, L77 (2000). 

\bibitem{Csahok:2000} Z. Csah\'ok, C. Misbah, F. Rioual, and A. Valance, Dynamics of aeolian sand ripples,
{\sl Eur. Phys. J. E} {\bf 3}, 71 (2000).

\bibitem{Gillet}
F. Gillet, Z. Csahok, and C. Misbah, Continuum nonlinear surface evolution equation for conserved step-bunching dynamics,
{\sl Phys. Rev. B} {\bf 63}, 241401 (2001).

\bibitem{FV}
T. Frisch and A. Verga, Effect of Step Stiffness and Diffusion Anisotropy on the Meandering of a Growing Vicinal Surface,
{\sl Phys. Rev. Lett.} {\bf 96}, 166104 (2006).

\bibitem{CKS} P. Politi and D. ben-Avraham, From the conserved Kuramoto-Sivashinsky equation to a coalescing 
particles model, {\sl Physica D} {\bf 238}, 156 (2009).

\bibitem{multiscale} R. Hoyle, {\sl Pattern formation} (Cambridge University Press, Cambridge, 2006).

\bibitem{Fredholm} D. Zwillinger, {\sl Handbook of differential equations} (Academic Press, San Diego, 1989)

\bibitem{nota}
If $q=q(X,T)$ when deriving the phase diffusion equation, at the end
(i.e., after taking derivatives) $q$ is a constant, $q=2\pi/\lambda$, because we are
interested in the perturbation of a periodic solution, where $q$ is constant.

\bibitem{Trefethen_book} L.~N. Trefethen, {\sl Spectral Methods in MATLAB} (SIAM, Philadelphia, 2000).

\bibitem{Boyd_book} J.~P. Boyd, {\sl Chebyshev and Fourier Spectral Methods} (Dover, New York, 2001).

\bibitem{Fornberg_book} B. Fornberg, {\sl A Practical Guide to Pseudospectral Methods}
(Cambridge University Press, Cambridge, 1996).

\bibitem{Canuto_book} C. Canuto, M.~Y. Hussaini, A. Quarteroni, and T.~A. Zhang,
{\sl Spectral Methods, Fundamentals in Single Domains} (Springer-Verlag, Berlin, 2006).

\bibitem{numerical_recipes} W.~H. Press, S.~A. Teukolsky, W.~T. Vetterling, and B.~P. Flannery,
{\sl Numerical Recipes, The Art of Scientific Computing}, 3rd Ed.\ (Cambridge University Press, New York, 2007).

\bibitem{Kassam:2005} A.-K. Kassam, and L.~N. Trefethen,
Fourth-Order Time-Stepping for Stiff PDEs, {\sl SIAM J. Sci. Comput.} {\bf 26}, 1214 (2005).

\bibitem{Cox:2001} S.~M. Cox, and P.~C. Matthwes,
Exponential Time Differencing for Stiff Systems, {\sl J. Comput. Phys.} {\bf 176}, 430 (2001).

\bibitem{Coleman:1994} T.~F. Coleman,  and Y. Li, On the Convergence of Reflective Newton Methods for 
Large-Scale Nonlinear Minimization Subject to Bounds, {\sl Math. Program.} {\bf 67}, 189 (1994).

\bibitem{Coleman:1996} T.~F. Coleman,  and Y. Li, An Interior, Trust Region Approach for Nonlinear Minimization Subject to Bounds,
 {\sl SIAM J.  Optimiz.} {\bf 6}, 418 (1996).

\bibitem{Villain_Guillot:08} S. Villain-Guillot, 1D Cahn-Hilliard dynamics: Ostwald ripening and application to modulated phase systems,
{\sl Phys. Lett. A} {\bf 372}, 7161 (2008).

\bibitem{alpha_sim} P. Politi and A. Torcini, Asymptotic and effective coarsening exponents in surface growth models, 
{\sl Eur. Phys. J. B} {\bf 53}, 401 (2006).

\bibitem{Peletier}
L.~A. Peletier and W.~C. Troy, {\it Spatial Patterns. Higher Order Models in Physics and Mechanics} (Birkh\"auser, Boston, 2001).  

\bibitem{2D}
C. Misbah and P. Politi, Phase instability and coarsening in two dimensions, {\sl Phys. Rev. E} {\bf 80}, R030106 (2009);
S. Biagi, C. Misbah and P. Politi, Coarsening Scenarios in Unstable Crystal Growth, {\sl Phys. Rev. Lett.} {\bf 109}, 096101 (2012).

\bibitem{Wolfram}
Eric W. Weisstein, {\it Jacobi Elliptic Function}. 
From MathWorld--A Wolfram Web Resource. http://mathworld.wolfram.com/JacobiEllipticFunctions.html 

\bibitem{strange_coarsening}
P.~C. Matthews and S.~M. Cox,
Pattern formation with a conservation law,
{\sl Nonlinearity} {\bf 13}, 1293 (2000).



\end{thebibliography}
\end{document}